\documentclass[fleqn,10pt]{wlscirep}
\usepackage[utf8]{inputenc}
\usepackage[T1]{fontenc}
\usepackage{graphicx}
\usepackage{subfig}

\newcommand{\kms} {$\mathrm{ km \; s^{-1}}\,$}
\newcommand{\msol} {M$_{\odot}$}

\newcommand{\rsol} {R$_{\odot}$}
\newcommand{\about} {$\sim$}

\newcommand{\ppxf}{{\tt ppxf}}

\title{End-to-end study of the home and genealogy of the first binary neutron star merger}

\author[1,*]{Heloise F. Stevance}
\author[1]{Jan J. Eldridge}
\author[2,]{Elizabeth R. Stanway} 
\author[2,]{Joe Lyman}
\author[3,4,]{Anna F. McLeod}
\author[2,5,]{Andrew J. Levan}

\affil[1]{The Department of Physics, The University of Auckland, Private Bag 92019, Auckland, New Zealand}
\affil[2]{Department of Physics, University of Warwick, Coventry, CV4 7AL, UK}
\affil[3]{Centre for Extragalactic Astronomy, Department of Physics, Durham University, South Road,  Durham DH1 3LE, UK}
\affil[4]{Institute for Computational Cosmology, Department of Physics, University of Durham, South Road, Durham DH1 3LE, UK}
\affil[5]{Department of Astrophysics/IMAPP, Radboud University, PO Box 9010,
6500 GL, The Netherlands}
\affil[*]{ hfstevance@gmail.com}
\begin{abstract}
\textbf{Binary neutron star mergers are one of the ultimate events of massive binary star evolution, and our understanding of their parent system is still in its infancy.
Upcoming gravitational wave detections, coupled with multi-wavelength follow-up observations, will allow us to study an increasing number of these events by characterising their neighbouring stellar populations and searching for their progenitors.
Stellar evolution simulations are essential to this work but they are also based on numerous assumptions. Additionally, the models used to study the host galaxies differ from those used to characterise the progenitors and are typically based on single star populations. 
Here we introduce a framework to perform an end-to-end analysis and deploy it to the first binary neutron star merger -- GW170817. 
With the Binary Population And Spectral Synthesis (BPASS) codes we are able to retrieve the physical properties of the host galaxy NGC 4993 as well as infer progenitor candidates.
In our simulations there is a $>$98\% chance that GW170817 originated from a stellar population with Z=0.010 born between 5 and 12.5 Gyrs ago.
By carefully weighing the stellar genealogies we find that GW170817 most likely came from a binary system born with a 13-24 \msol primary and 10-12 \msol secondary which underwent two or three common envelope events over their lifetime.}
\end{abstract}
\begin{document}

\flushbottom
\maketitle
% * <john.hammersley@gmail.com> 2015-02-09T12:07:31.197Z:
%
%  Click the title above to edit the author information and abstract
%
\thispagestyle{empty}
The gravitational-wave event GW170817 was associated with an optical transient AT~2017gfo  which allowed to identify the host: a nearby Sd0 galaxy (NGC~4993) roughly 40 Mpc away. 
To infer the progenitors of the binary neutron star merger we first need characterise the star formation history (SFH) and metallicity content of the host. 
This information is encoded in the Spectral Energy Distribution (SED) of the galaxy and estimates can be retrieved from fitting the observed SEDs with model SEDs whose composition and mix of ages is known. 
The ages and metallicities derived at this step can be used in conjunction with the neutron star masses derived from the gravitational wave event by LIGO/Virgo to identify progenitor candidates in stellar population models.  
Once good matches have been found their evolutionary route can be traced back and we can look for patterns in how these different channels are correlated to initial total masses, periods, and mass ratios.
The full extent of this process therefore requires stellar modeling, population synthesis and spectral synthesis -- historically this has required the use of different models at different steps. 

Using the data products from BPASSv2.2.1\cite{eldridge2017, stanway2018} the entire process can be followed end-to-end. There are three key aspects to the BPASS simulations: (i) detailed binary stellar evolution models calculated with a custom version of the Cambridge STARS code\cite{eggleton1971,eldridge2008,eldridge2017}; 
(ii) population synthesis\cite{eldridge2017}, where each stellar population is the result of a starburst from 10$^6$ \msol\, of material at a given metallicity Z; (iii) spectral synthesis\cite{stanway2018}, such that each stellar model includes predicted absolute photometry and each 10$^6$ \msol\, population has an associated Spectral Energy Distribution (SED).
Every star system in a given population will have an associated number density (how many such systems are expect to occur per 10$^6$ \msol) that is dependent on the initial mass function  and initial binary parameter distribution\cite{moe2017}. 
In this work we use results from the fiducial initial mass function, which is a Kroupa\cite{kroupa2001} prescription with maximum mass of 300 \msol.
The main difference between BPASS and other codes typically used to study binary neutron star mergers\cite{belczynski2002, kruckow2018, COMPAS, marchant2021} is that our simulations are not run to specifically fit compact object populations.
Instead, BPASS provides a large library of predictions that self-consistently reproduce key observables of massive star and transient populations \cite{eldridge2013, steidel2016, bestenlehner2020, stevance2021, runco2021, stevance2022, briel2022}, albeit on a coarser grid of initial conditions.

The model data used in this study had been previously published but extensive data pipelines had to be created.
The python package {\tt hoki}\cite{stevance2020} is the main interface to the BPASS models and new features have been implemented for this study: the release of {\tt hoki v1.7} associated with this publication makes all these pipelines openly available (see Code Availability) to allow other teams to follow this framework on future transient studies and for future kilonova follow-up in the coming years.

%%%%%%%%%%%% SED FITTING SED FITTING SED FITTING
\section*{A binary view of NGC 4993}
\begin{figure}
    \centering
    \includegraphics[width=11cm]{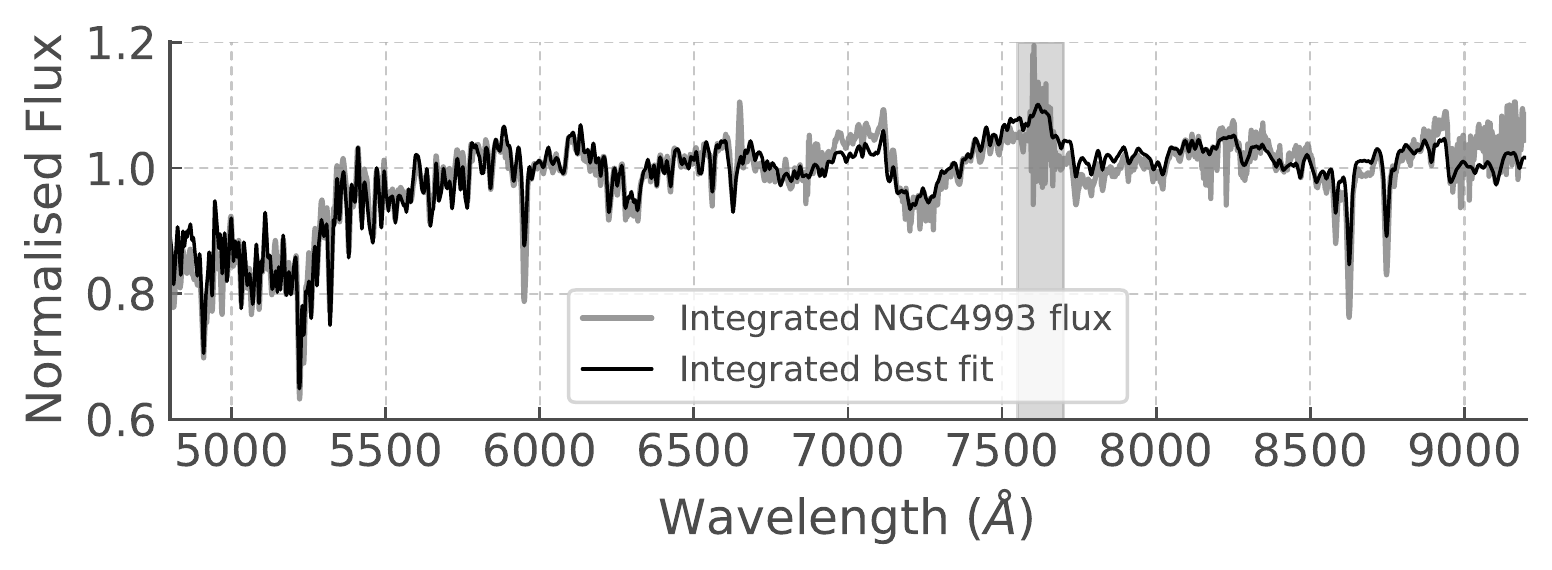}
    \includegraphics[width=11cm]{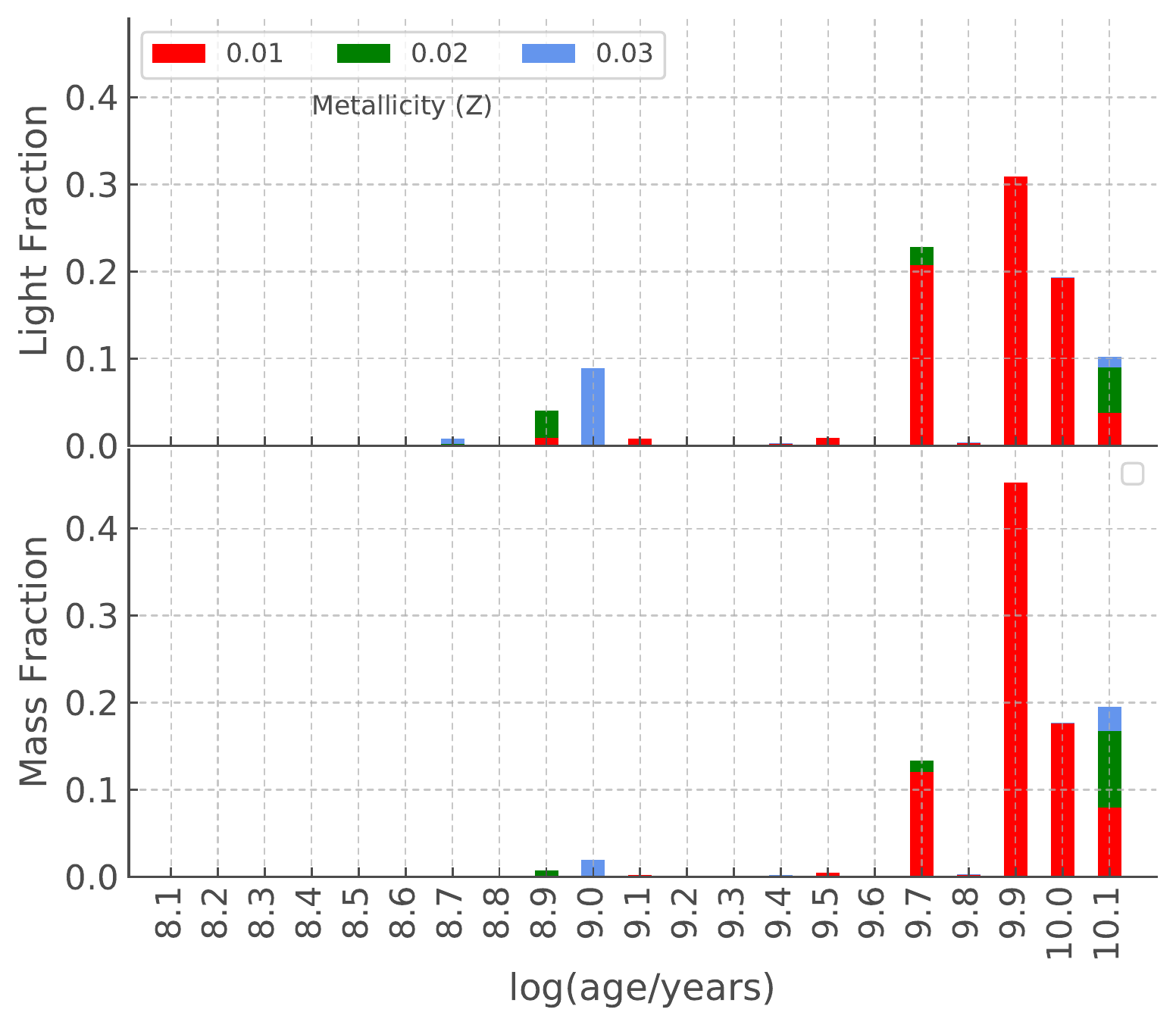}
    \caption{SED fit and resulting SFH of NGCC~4993. \textbf{Top Panel:} Integrated flux of NGC~4993 and the integrated best fit over all voronoi bins. \textbf{Bottom Panel:} SFH in terms of light fraction (middle panel) and mass fraction. The mass fraction is calculated by comparing the flux of templates required to fit each bin to the flux predicted for BPASS models created with 10$^6$\msol\, at the zero-age main sequence and accounting for the stellar mass remaining given the age bin.  The grey shaded region highlights a zone contaminated by strong telluric lines and the absorption feature at 6795\AA\, is also a telluric remnant.}
    \label{fig:ngc4993}
\end{figure}

The best data set of NGC~4993 for this study is the MUSE Integral Field data obtained on 2017 August 18\cite{levan2017} as a follow-up from the detection of the kilonova AT~2017gfo.
The data cube allows us to perform a spatially resolved fit of the galaxy (see Supplementary Information SI.1) using our custom SED templates (see Methods).
The advantage of integral field data is that thanks to its large field-of-view, it has the potential to reveal spatial patterns in the star formation history or metallicity distribution of a given galaxy - in this particular case, however, we find that the distribution of the populations in NGC~4993 is mostly homogeneous. 
Had this not been the case, it would have been necessary to consider the likely significant offset (\about a few kpc\cite{fong2010}) between the birth of the massive stars and merger location of their remnants, which is expected as a result from supernova kicks. 
For the purposes of this analysis we can therefore focus on the integrated flux of NGC~4993 and the integrated best fit and its resulting star formation history (see Figure \ref{fig:ngc4993}).
Overall NGC~4993 has two main components: a) 95\% of the mass is made of an old and metal-poor (Z=0.010) population with age \about log(age/years) $\ge$ 9.7 ($\ge$ 5 Gyrs) and a peak at log(age/years) \about 9.9 (\about 8 Gyrs) corresponding to the peak of star formation in the Universe\cite{madau2014}; b) $\lesssim$ 5\% of the mass is composed of a younger stellar population (\about log(age/years)=8.9-9.0, or 790 Myrs to 1Gyrs) with solar to super-solar metallicity (Z=0.020, 0.030). 

The solar and super-solar components at log(age/years)=10.1 have been inferred to be spurious after investigation (see Supplementary Information SI.1) and they are therefore not taken into account in our search criteria. 
Additionally, given the complex uncertainties associated with SED fitting and stellar modeling, it would not be reasonable to conclude that the Z=0.020 and Z=0.030 components are separate populations - it is more likely that the younger episode of star formation was not instantaneous (compared to the size of our age bins) and that the metallicities of the stars created over that period have a range spanning the solar to super-solar regime.

Previous SED fitting of NGC~4993 by independent groups covered a broad range of metallicities\cite{blanchard2017, pan2017, levan2017,im2017} (from 20\% solar to 2.5 times solar).
The most straightforward comparison we can make is with the results from a study that used the same data set we are utilising in this work: their best fit SED models include solar and super-solar stellar populations, as well as a juvenile component \about 10 Myrs old\cite{levan2017}, which we do not recover at all. 
We find that the discrepancy is a result of the differences between the stellar models and atmosphere models used in the previous study\cite{bc03} and the ones used in BPASS (see SI.1.4). 
The key conclusion is that different stellar models can yield different quantitative results even if they tell the same qualitative story.
That is because every model uses a unique set of assumptions and prescriptions to approximate the complex physics of stellar populations, stars, and stellar atmospheres.
This is one of the key motivations for using BPASS end-to-end -- to both determine the properties of the host environment and investigate the genealogy of the transient: our set of assumptions are homogeneous throughout our analysis.

%%%%%% GENEALOGIES GENEALOGIES GENEALOGIES
\section*{The possible genealogies of GW170817 in BPASS}

\begin{figure}[h!]
    \centering
    \includegraphics[width=17cm]{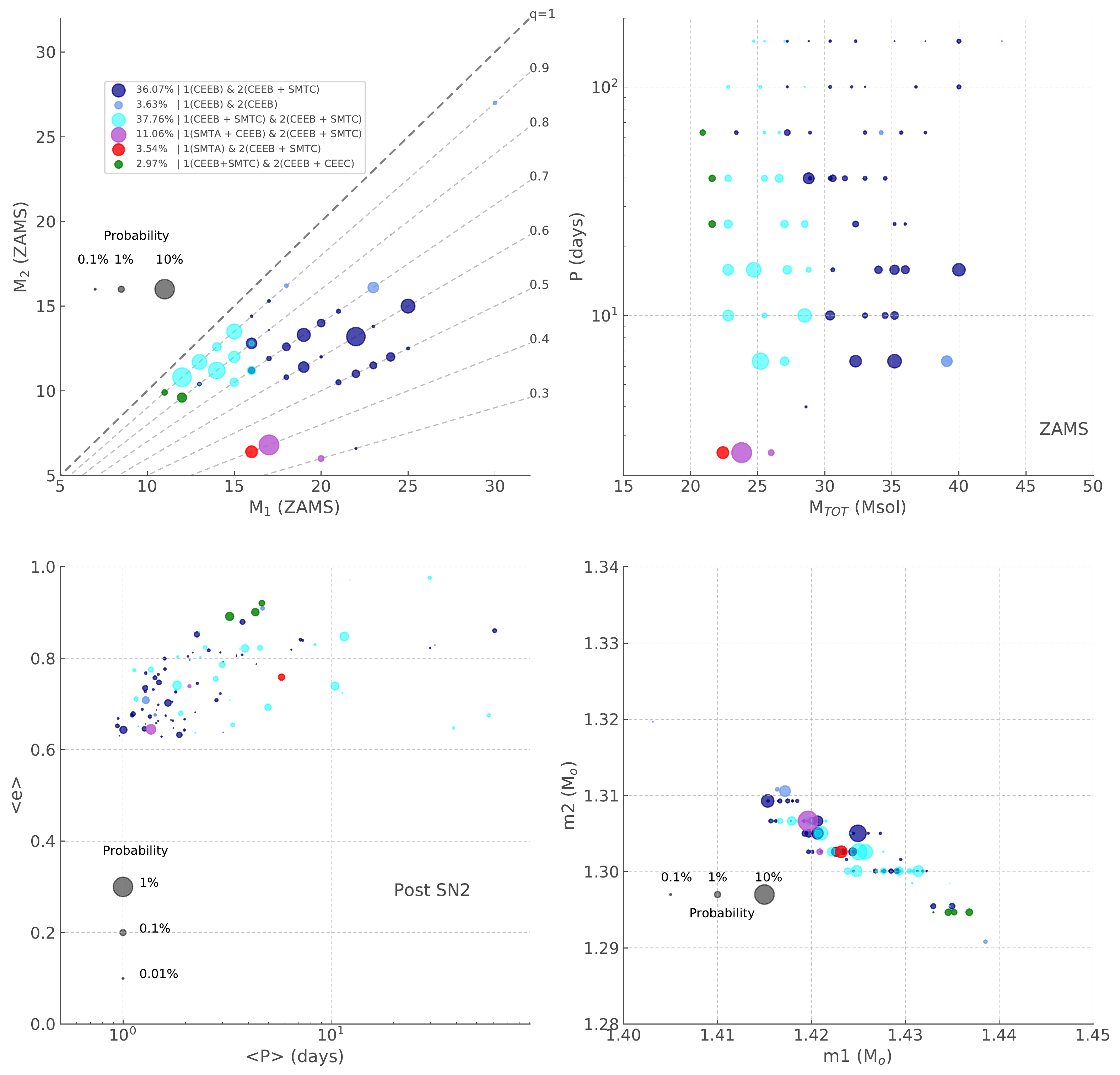}
    \caption{Summary characteristics of the progenitor candidates of GW170817 split by evolutionary channels. The probabilities correspond to the total weights (see Methods) in percent. Blue hues correspond to systems where both the primary and secondary undergo common envelope evolution (CEE), red hues to systems where the primary undergoes stable mass transfer (SMT) on the main sequence (case A) and green is used for systems where the secondary undergoes two phases of CEE, one during helium burning (case B) and one post helium burning (case C). The codes in the legend describe the key evolutionary phase of the systems, for example channel ``1(CEEB) \& 2(CEEB+SMTC)" is a system where the primary star (1) underwent case B CEE, died, then the secondary (2) underwent case B CEE and  subsequently case C SMT. Only the evolutionary channels where the summed weight is greater than 0.05\% are shown.}
    \label{fig:ap4_lowspin}
\end{figure}

The BPASS stellar models that match the properties of the stellar population of NGC~4993 and the neutron star masses inferred from the gravitational wave event GW170817 are identified and weighted (see Method and Supplementary Information).  
We find that Z=0.010 is 99\% of the weights (or probability), so the following discussion will be entirely focused on models of this metallicity. 
We present in Figure \ref{fig:ap4_lowspin} graphs summarising key characteristics of the dominant evolutionary channels that lead to the progenitors of GW170817. 
Most noticeably, common envelope evolution (CEE) -- where a star expands beyond the second Langrangian point and engulfs its companion\cite{paczynski1976}-- is ubiquitous in our systems.
CEE has long been established as an important step towards the creation of binary neutron stars\cite{bhattacharya1991, tauris2006}, however commonly invoked channels in the literature only exhibit one episode of CEE\cite{tauris2017, vignagomez2018}. 

The majority of our progenitor candidates, however, undergo CEE twice over the life of the system (blue hues in Figure \ref{fig:ap4_lowspin}): once from the primary star leaving the main sequence, then a second time when the secondary star leaves the main sequence (case B CEE). 
In these channels the primary star has Zero Age Main Sequence (ZAMS) mass $M_1 \ge$ 12\msol and the mass ratio (q:=$M_2 / M_1$) covers a broad range of parameter space from 0.5 to 0.9 (which is the maximum q included in the grid of initial parameters in the BPASS models). 

There is however a distinct split within this group of systems, between those where the primary stars undergo CEE with no further binary interactions, and those exhibiting stable mass transfer (SMT) later in life (case C SMT) -- in Figure \ref{fig:ap4_lowspin} the cyan markers.
These models dominate the high mass ratio region of parameter space (q$\ge0.7$) and they only occur for primary star with ZAMS mass $M_1 \le$ 16\msol, leading to total masses at birth $\le$30\msol.
On the other hand, the genealogies where the primaries do not undergo a second episode of binary interaction prefer lower mass ratios (0.5--0.8) and  $M_1 \ge$ 16\msol, for initial total masses greater than 27\msol. 
The distinction between these systems is not as obvious when looking at the final properties of the binaries: the periods and eccentricities after the second supernovae and the final masses of the neutron stars $m_1$ and $m_2$ (note that these refer to the masses of the most and least massive neutron star, respectively, which does not necessary correspond to the masses of the neutron stars from the primary and secondary as mass inversion is common). 
Although it would appear that the cyan markers and blue markers may occupy slightly different regions in the final period-eccentricity plot (bottom left panel of Figure \ref{fig:ap4_lowspin}), this is not a consistent feature when using alternative neutron star equations of state or considering a high spin prior for the inferred $m_1-m_2$ from the LIGO/Virgo detection (see Supplementary Information), and there is no clear distinction in the range of binary neutron star systems they produce. 
It is worth noting here that the range of allowed neutron star masses in the BPASS models presented here does not extend across the full allowed range of $m_1-m_2$ from LIGO/Virgo, for example we find no neutron stars with masses below 1.25\msol: this is for the most part due to the fact that the baryonic neutron star mass from a supernova explosion is taken to be 1.4 at minimum. 
This assumption was initially made to limit the number of detailed models that we needed to calculate; lower baryonic masses may be possible from electron-capture supernovae or through accretion-induced collapse but these are not accounted for in this work. After the second supernova however we use the remnant mass estimated from determining how much mass is left after the stellar envelope is lost for 10$^{51}$ ergs of energy.
This does not impact how we compare our models to each other but is important to remember if comparing to other predictions. 

Now we turn our attention to another distinct type of progenitor system retrieved in our searches, that are characterised by primary stars which undergo stable mass transfer on the main sequence (red and purple markers in Figure \ref{fig:ap4_lowspin}). 
They are very distinct to other scenarios in two ways: their initial mass ratio is always low (q$\le$0.5) and their initial periods are short (2.5 days in Figure \ref{fig:ap4_lowspin}, which is the lowest initial period on the BPASS grid of initial parameters). 
The short periods are immediately correlated with the fact that these systems undergo mass transfer early on, while the consistently low mass ratios are a result of the necessary condition for this main sequence mass transfer to be stable, as it would otherwise lead to mergers and remove the systems from the pool of viable binary neutron star progenitor candidates. 
Once again we see that the final properties (periods, eccentricities and neutron star masses) are found in similar areas of parameter space as the systems originating for the other channels (also see Supp. Figs. 10, 11, 12). 
This is a clear demonstration that binary systems with very different initial properties (total mass, mass ratio, initial periods) can lead to very similar binary neutron star systems as a result of the binary interactions that they undergo over the course of their evolution. 

Finally there is one more type of evolutionary route seen in Figure \ref{fig:ap4_lowspin}: the systems where the secondary star undergoes two distinct epochs of CEE (green markers). 
These are found to have the lowest total mass of our progenitor candidates but most notably they always result in high final eccentricity and periods following the second supernova (e$>0.85$),  and periods of several days; they also result in binary neutron star systems with lower $m_2$ and higher $m_1$ than the majority of our other matches. They are the only genealogies to somewhat distinguish themselves from the others in their final properties. 

\begin{figure}[h!]
    \centering
    \includegraphics[width=12cm]{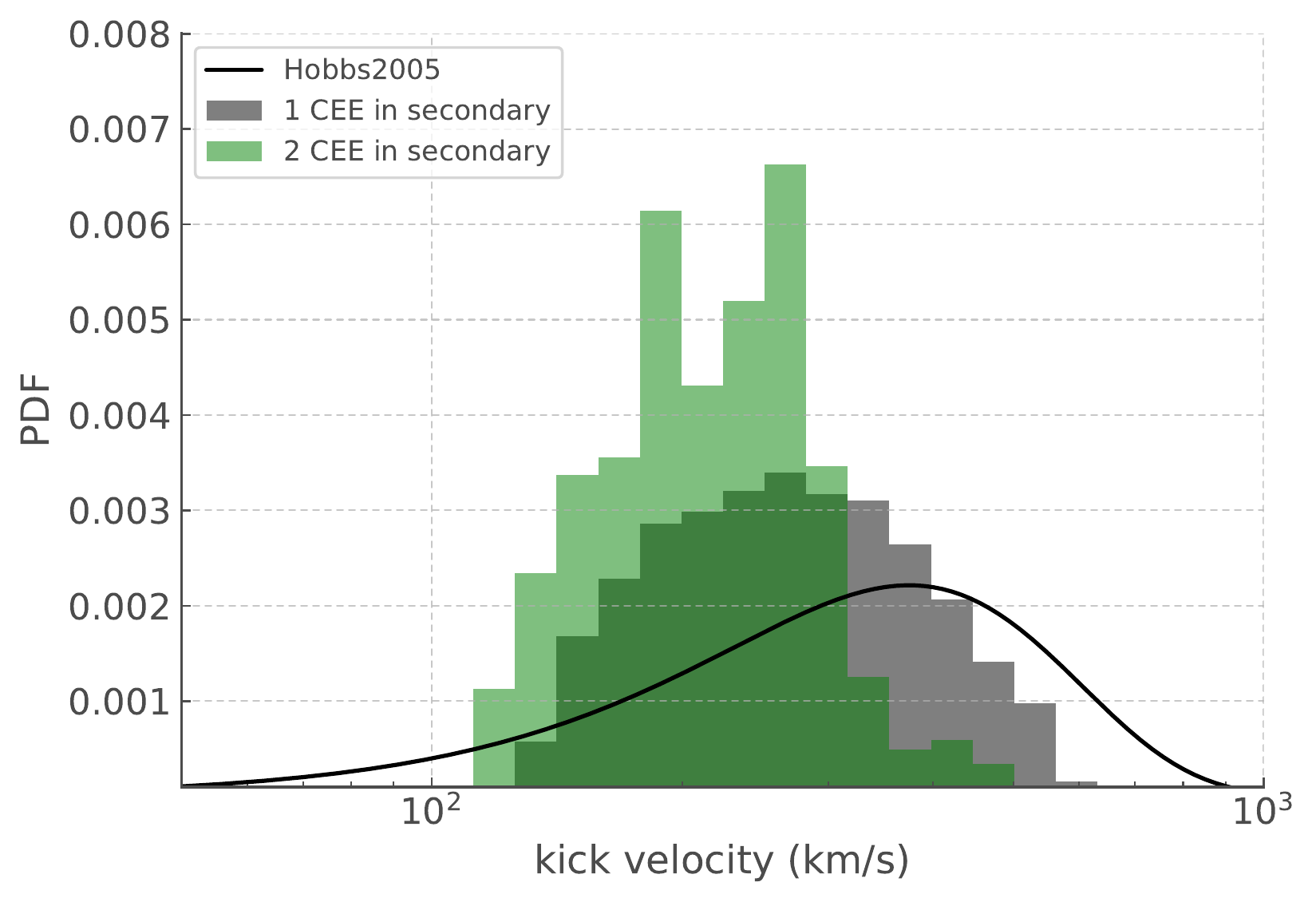}
    \caption{Kick velocity distributions of GW170817 progenitor channels. The kick velocities are grouped according to whether the secondary star undergoes one or two episodes of CEE (sampling size 11035 and 1143 data points, respectively). Note that this histogram includes data for the AP4 and WFF1 EoS (see SI.3).}
    \label{fig:kicks}
\end{figure}

Another key distinction between the systems where the secondaries undergo two CEE and those who undergo only one is their kick velocity distributions (see Figure \ref{fig:kicks}).
Both distributions are skewed towards lower velocities than the Hobbs distribution we used for sampling, which is expected as the Hobbs distribution is empirically determined from runaway pulsars -- it is therefore reasonable for neutron stars that remain bound to require lower kicks. 
What is notable is that the secondaries that go through a second CEE prefer lower velocities than the other systems, and velocities greater than \about 350--400 \kms are disfavoured or prohibited. 
The need for lower kicks as well as the consistently high eccentricities exhibited by the binary neutron stars from these genealogies suggest that the supernova kick needs to have a specific direction and amplitude for these systems to provide matching progenitors. 
We find that this is a result of their final separation before the second supernova being consistently higher (tens of solar radii) than that of the other genealogies.

Overall these systems are initially not typically suited to forming binary neutron star that could merge within a Hubble time and create an event like GW170817, as they maintain a very high separation following the first episode of CEE ($>$ 100 \rsol -- see e.g. Sup. Fig. 13). But the occurrence of a second episode of unstable mass transfer during carbon core burning results in further orbital shrinkage. Still their orbit is quite wide and a kick that is too strong or with the wrong orientation leads to the binary becoming unbound. If the kick has a fortuitous velocity the system can remain bound and the high eccentricity allows for a short enough in-spiral time to match our star formation history.

The fact that our secondaries with two episode of CEE consistently have high separations is particularly interesting because such evolutionary channels have previously been presented in the literature \cite{belczynski2018, chruslinska2018} but lead to extremely tight orbits (semi-major axis \about 0.1\rsol).
This illustrates how differences in how the CEE phase is modelled (how much the orbit of the system shrinks as a result and how much envelope is lost or retained in the process) can result in stark changes in the products of binary stellar evolution. 
One of the main sources of divergence is that our stellar evolution code evolves the stellar interior and envelope in detail as opposed to rapid evolution codes (see discussion in SI2.4) which require parameterisations to account for the response of the stellar envelope. 
This does not mean that BPASS treats CEE better -- as a 1D hydro-static stellar evolution model it cannot accurately simulate the dynamical phase of CEE and our assumptions need to be tested on a large scale to quantify the effects of energy conservation, the core-envelope boundary and the conditions of onset of unstable mass transfer. 

Another important point to consider is that in this study we have only considered isolated binary evolution, but there are other avenues to creating compact objects and binary neutron stars, such as dynamical interactions or nuclear cluster formation\cite{mandel2022}.
In the case of GW170817 it is appropriate to focus on isolated binary formation as other routes have been found to be highly unlikely by previous groups. 
The rates of dynamical interaction induced binary neutron star mergers is expected to be low in the local universe\cite{ye2020}; furthermore, a study comparing the isolated binary route (using {\tt StarTrack}\cite{belczynski2002,belczynski2008}), dynamical interaction and nuclear cluster formation has found that the latter two have rates 3 orders of magnitude lower than the isolated binary interaction scenario\cite{belczynski2018}.
Although their stellar evolution code and ours are distinct, the local rates for binary neutron star mergers that they retrieve are \about 10 to 20 times lower than our estimates from combining BPASS with cosmological simulations\cite{briel2022} - therefore our binary neutron star merger rates would still dominate compare to their estimates of alternative routes.
It is also worth noting that the most up-to-date results on binary neutron star merger rates are consistent between BPASS and {\tt StarTrack}\cite{olejak2021}.

With the new observing run of gravitational wave detectors expected to start in 2023, it is anticipated that \about10 new binary neutron star mergers could be discovered in the span of 12 months.
With a decrease in the search sky area (by a factor of 8 on average)\cite{nguyen2021} and the increased follow-up capability of optical facilities, it is highly likely that the next year will provide new data sets of direct observations of binary neutron star mergers and their host galaxies. 
The deployment of our end-to-end pipeline will allow us to characterise the neighbouring stars and quantify the most likely genealogies specific to different binary neutron star mergers. 
As we have seen in the case of GW170817, many different routes can lead to suitable matches to a given transient, but as the sample of known kilonovae grows, the model matching and weighing methods presented could reveal clear trends in their progenitor systems.
By being able to use the same stellar evolution simulation across our entire analysis, we know exactly what assumptions are involved and that they remain homogeneous throughout. 
This is important for self-consistency within our studies but is also relevant for future analyses that will be undergone in the next decade as the sample of binary neutron star mergers grows; as our collective understanding of complex evolutionary phases (such as CEE) evolves it will be easier to systematically assess the effects of our current assumptions and build a robust library of progenitor candidate models.

%%%%%%%%%%%%%%%%%%%%%%%%%
%%%  METHODS METHODS %%%%
%%%%%%%%%%%%%%%%%%%%%%%%%

\section*{Methods}
Throughout this study with use  BPASSv2.2.1 (see Data Availability) which includes a library of over 250,000 stellar models computed over 13 metallicities and contains detailed (i.e. the stellar interior is computed) evolution for both single and binary systems\cite{eldridge2008, eldridge2017, stanway2018}.
In BPASS each binary system has a unique combination of initial masses (Zero-Age Main Sequence -- ZAMS), period and metallicity. 
The population synthesis provides a number $N_{\rm}(M_{1,\rm zams}, M_{2,\rm zams}, P_{\rm zams})$, which quantifies how many times each system is expected to occur in a 10$^6$\,\msol\, population for a given initial mass function. 
The properties of each stellar population  in most of the data products (e.g. population SEDs) is binned to 51 time intervals in log(age/years) from 6 to 11.0 in 0.1 dex increments, but it is worth noting that in the stellar models time steps are recorded with variable interval such that the evolution of the star is sufficiently well sampled.
Subsequently we can perform spectral synthesis: the temperature, luminosity, surface gravity and surface composition of each surviving star in each time step is used to identify a matching stellar template from an extensive library that combines publicly available template spectra of main sequence stars, giants, Wolf-Rayet stars, white dwarfs and other stars\cite{stanway2018}. These spectra are combined using the population synthesis occurance rate $N_{\rm}(M_{1,\rm zams}, M_{2,\rm zams}, P_{\rm zams})$ to determine the composite SED that would be observed from the unresolved stellar population. 
Spectra are generated on a grid with a uniform interval of 1\AA\, in wavelength, spanning from the far ultraviolet to the mid-infrared.

\subsection*{SED fitting}
To fit BPASS synthetic spectra to the observations we use the Penalized Pixel-Fitting (\ppxf) algorithm\cite{cappellari2017}.
Pre-processing is required to make BPASS templates compatible with \ppxf. Since it is dependent on the observational data to be fitted,  we developed {\tt hoki}\cite{stevance2020} features to create the custom template SEDs and handle the \ppxf\, outputs. 
The BPASS SEDs contain 100,000 wavelengths and are calculated for 51 ages in 13 metallicities. 
We restrict the wavelength range to that of the observational data with a buffer of +/-50 \AA, and we constrain our templates to the 42 BPASS ages within a Hubble time: log(age/years)=6.0 to 10.1 in 0.1 increments.
Additionally the results presented here were obtained from templates spanning only 3 metallicities: Z=0.010, 0.020 (solar metallicity) and 0.030. 
Due to the fact that SED fitting is by nature an ill-conditioned inverse problem -- meaning that it suffers from severe (and unsolvable) degeneracies \cite{cappellari2017} -- and that \ppxf\, is a mathematical optimiser which has no way of evaluating how physically meaningful the results are, providing too many templates can lead to uninformative or unphysical best fits.
Consequently we narrow down parameter space to the 3 metallicities that were found to be the most informative and be in line with previous estimates for the metallicity of NGC~4993 reported in the literature\cite{blanchard2017,pan2017,im2017,levan2017}.
Extensive discussion of this process can be found in the Supplementary information (SI.1.3) and the full data analysis code used to perform the fits can be found as part of the data release associated with this publication (see Data Availability). 
It is also worth noting that because of the nature of the problem the solution presented here is one of potentially several best solutions - therefore the values of the light/mass fraction shown in Figure \ref{fig:ngc4993} should be taken as qualitative indicators not precise quantities (doing so may introduce unwanted systematics to the analysis). 

\subsection*{Matching models and weighing genealogies}
In order to identify all suitable progenitor candidates for GW170817 in the BPASS predictions we match 3 key quantities: the metallicity, the delay time (time between the Zero-Age Main Sequence -- ZAMS --
and the merger), and the mass of the neutron stars $m_1$ and $m_2$.
The metallicity dependent ages of the two populations described earlier can be used directly to constrain our models.
Then matching the neutron star masses requires comparing to the posterior distribution found by the LIGO/Virgo collaboration\cite{abbott2019}. In the main text we show results obtained when matching the $(m_1,m_2)$ posterior found using the low spin priors, but results for the high spin priors can be found in the Supplementary Information (SI.3). 
A further complication arises when we consider that predicted neutron star masses in stellar evolution codes are baryonic masses, whereas the neutron star masses reported from observed data are gravitational masses. 
The conversion between the two is dependent on the neutron star equation-of-state (EoS), and although EoS independent conversions exist, they are subject to potentially large systematic uncertainties (up to \about 0.1 \msol)\cite{gao2020}.
When converting from baryonic to gravitational mass, the choice of EoS will impact the individual neutron star masses as well as the mass ratio, leading to different weighting for each suitable progenitor channels. 
We chose to perform the mass conversion (and subsequent genealogy exploration) for the three EoS which show the best consistency with the mass and areal radius posterior distribution found by LIGO/Virgo from  GW170817\cite{abbott2018} (WFF1\cite{WFF1}, AP4\cite{AP4} and MPA1\cite{MPA1} -- see SI2.2).
In the main text we presented the results obtained for the AP4 EoS, and SI.3 presents results obtained for all three considered EoS. 

In addition to matching the observed properties to our models we can also weigh each genealogy by taking into account the effects of the initial mass function and the effects of natal kicks imparted by supernovae. 
The former are already given by the BPASS models while the latter requires supplementary numerical simulations: for each supernova event we performed 2000 kicks by randomly sampling a Maxwellian distribution with $\sigma=265$\kms\cite{hobbs2005} for the speed and a uniformly isotropic distribution for the direction of the supernova kicks.  
The final weight of each genealogy accounts for the initial mass function, the supernova kicks, and the $(m_1,m_2)$ posterior distribution: An extensive description of the matching and weighing procedure can be found in SI.2, and each step can be reproduced using the jupyter notebooks included in the data release.

\section*{Data Availability}
BPASSv2.2.1 can be found at \url{https://bpass.auckland.ac.nz} or \url{https://warwick.ac.uk/bpass}. The LIGO/Virgo posterior distributions are publically avialable from \url{https://dcc.ligo.org/LIGO-P1800370/public}. 
Finally the data resulting from the numerical simulations of supernova kicks used to perform this analysis (main text and supplementary information) are main publicly available at \url{https://doi.org/10.5281/zenodo.7138935}

\section*{Code Availability}
The hoki python package is available on GitHub \url{https://github.com/HeloiseS/hoki} and published in the Journal of Open Source Software \url{https://doi.org/10.21105/joss.01987}. 
The data analysis code used to perform this analysis (main text and supplementary information) are main publicly available at \url{https://doi.org/10.5281/zenodo.7138935}.

\section*{Acknowledgements}
HFS and JJE acknowledge the support of the Marsden Fund Council managed through Royal Society Te Aparangi.
HFS is thankful to Lorenza Della Bruna, Angela Adamo and Chris Usher for private communications regarding the voronoi binning and ppxf algorithms.
HFS is grateful to Stephen Smartt for his comments on the final draft of the manuscript. 
ERS acknowledges funding from the UK Science and Technology Facilities Council (STFC) through Consolidated Grant ST/T000406/1.
AJL has received funding from the European Research Council (ERC) under the European Union’s Horizon 2020 research and innovation programme (Grant agreement No.725246).

\section*{Author contributions statement}

H.F.S. is the lead developer of hoki, created the pipelines to make BPASS SED templates, refactored the TUI codebase, performed the SED fitting, and led the writing of this manuscript. 
J.J.E. conceived the project, is one of the lead developers of BPASS, wrote TUI, advised throughout the analysis and writing process, and contributed text to the supplementary information.
E.S. is one of the lead developers of BPASS and commented on early and later drafts of the manuscript.
J.L. provided relevant expertise on the SED fitting procedure and analysis of NGC~4993 as well as the data reduction used for this paper.
A.F.M. advised on the handling of MUSE data and SED fitting.
A.J.L. led initial acquisition of the data, provided text and advised on the manuscript.

\section*{Additional information}
\textbf{Competing interests}: There are no competing interests to report. 

\section*{Figures}

\end{document}

\maketitle

This is the supplementary information for our study of the host environment and progenitor candidates of GW170817 with BPASS. We used MUSE integral field data of NGC4993 obtained on 2017 August 18 previously presented in the literature\cite{levan2017}.
In Section \ref{sec:SEDfit} we provide further detail of our new voronoi binning for these data, our assessment of the best choice of SED fitting parameters, and we show a comparison to the results obtained previously with different SED templates.                                                                                                                                                                                                                                                                                                            
In Section \ref{sec:BPASS} we give an extended description of our BPASS models and auxiliary numerical simulations; we also detail how our genealogies are weighted and provide a thorough discussion of Common Envelope Evolution within our simulations. 
Finally in Section \ref{sec:results} we present the full range of our resulting progenitor genealogies for three considered neutron star equations of state, and compare to the evolutionary channels expected from the literature. 

%%%%%%%%%%%%%%%%%%%%%%%%%%%%%%%%%%%%%%%%%%
%%%%%%%%%%%%%%%%%%%%%%%%%%%%%%%%%%%%%%%%%%
%  SED FITTING  SED FITTING  SED FITTIMG
%%%%%%%%%%%%%%%%%%%%%%%%%%%%%%%%%%%%%%%%%%
%%%%%%%%%%%%%%%%%%%%%%%%%%%%%%%%%%%%%%%%%%
\section{SED fitting of the host environment}
\label{sec:SEDfit}

%%%%%%%%%%%%%%%%%%%%%
% VORONOI AND SNR
%%%%%%%%%%%%%%%%%%%%%%
\subsection{Voronoi binning}
\begin{figure}
    \centering
    \includegraphics[width=17cm]{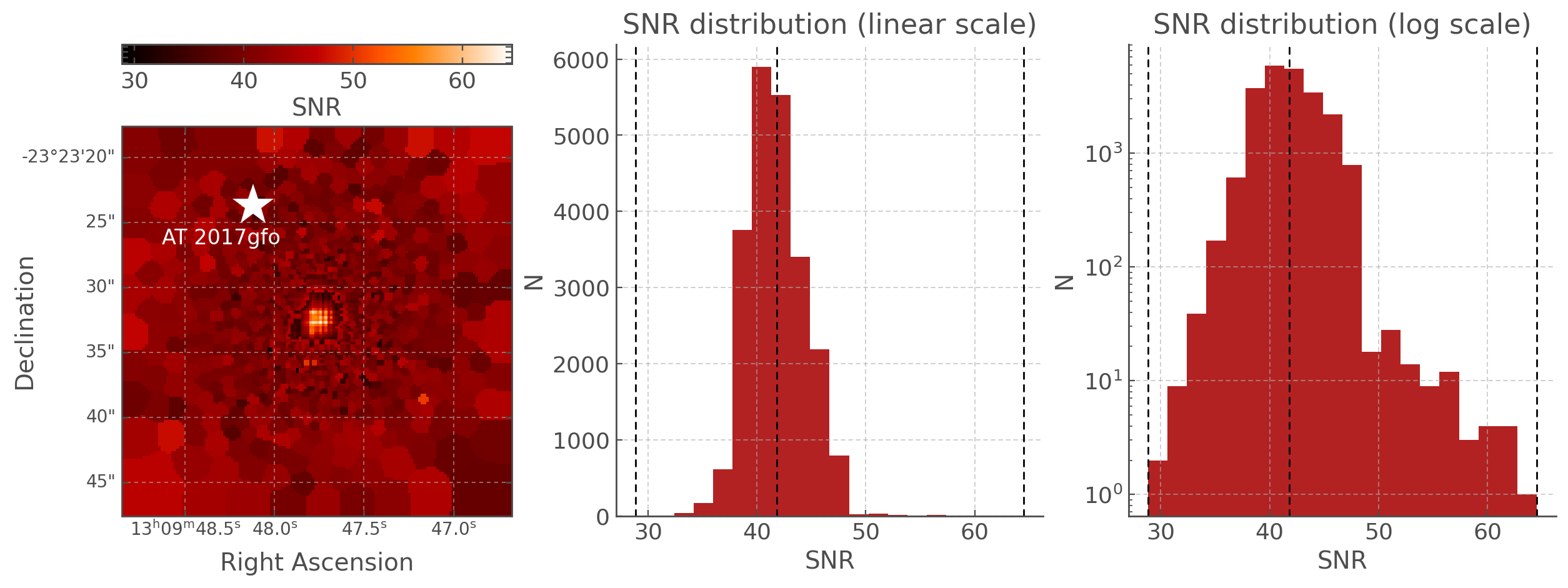}
    \caption{Signal-to-Noise ratio (SNR) resulting from voronoi binning of NGC~4993. The SNR of each voronoi bin (left panel) and SNR distribution (middle and right panels). The dashed line in the SNR histograms show the minimum, mean and maximum SNR values.}
    \label{fig:snr_voronoi}
\end{figure}

We performed new Voronoi binning using the {\tt vorbin} code\cite{cappellari2003} with a target Signal-to-noise ratio (SNR) of 40 - meaning that the algorithm will optimise the binning of adjacent pixels to homogenise the SNR across the image to be around 40, within some stochastic noise (for a review on vornoi binning see\cite{cappellari2009}).
The binning algorithm is provided with a pixel-wise SNR which we calculate by taking the average SNR between 5590 and 5680 \AA, as this region of the spectrum is devoid of strong lines and is representative of most of the spectrum. 
We retrieve 1494 voronoi bins (see left panel of Sup. Fig. \ref{fig:snr_voronoi}) with a mean SNR of 41 and 90\% of the SNR distribution between 38 and 46.

We note that the central region of NGC4993 shows a high SNR region (\about 60 -- left panel of  Sup. Fig. \ref{fig:snr_voronoi}) surrounded by a ring-like structure of low SNR around 35. 
This is expected as a result of the transition from a high flux region (nucleus of NGC4993) to the lower flux regions (outskirts of the galaxy): the pixels in the central regions of the galaxy are sufficiently bright to meet the SNR criteria without the need for binning. 
The lowest SNR region are pixels that individually approach the SNR criteria but would surpass it if they were binned together.
Further out of these region, adjacent pixels start forming voronoi bins and the SNR distribution becomes homogenised. The only consequence of this effect on the SED fitting is the resulting reduced $\chi^2$ (see Sup. Fig. \ref{fig:chi2_clean_comp} - see right panel) and it does not affect our results. 
Note that generally such effects can affect SED fitting results if the SNR target is too low and the binned spectra remain too noisy - so user caution is advised when deploying this method. 

%%%%%%%%%%%%%%%%%%%%%%
% PARAMETERS AND LIERS
%%%%%%%%%%%%%%%%%%%%%%%
\subsection{Best fit parameters, Low Ionisation Emission Region (LIER) and reddening}
\begin{figure}
    \centering
    \includegraphics[width=17cm]{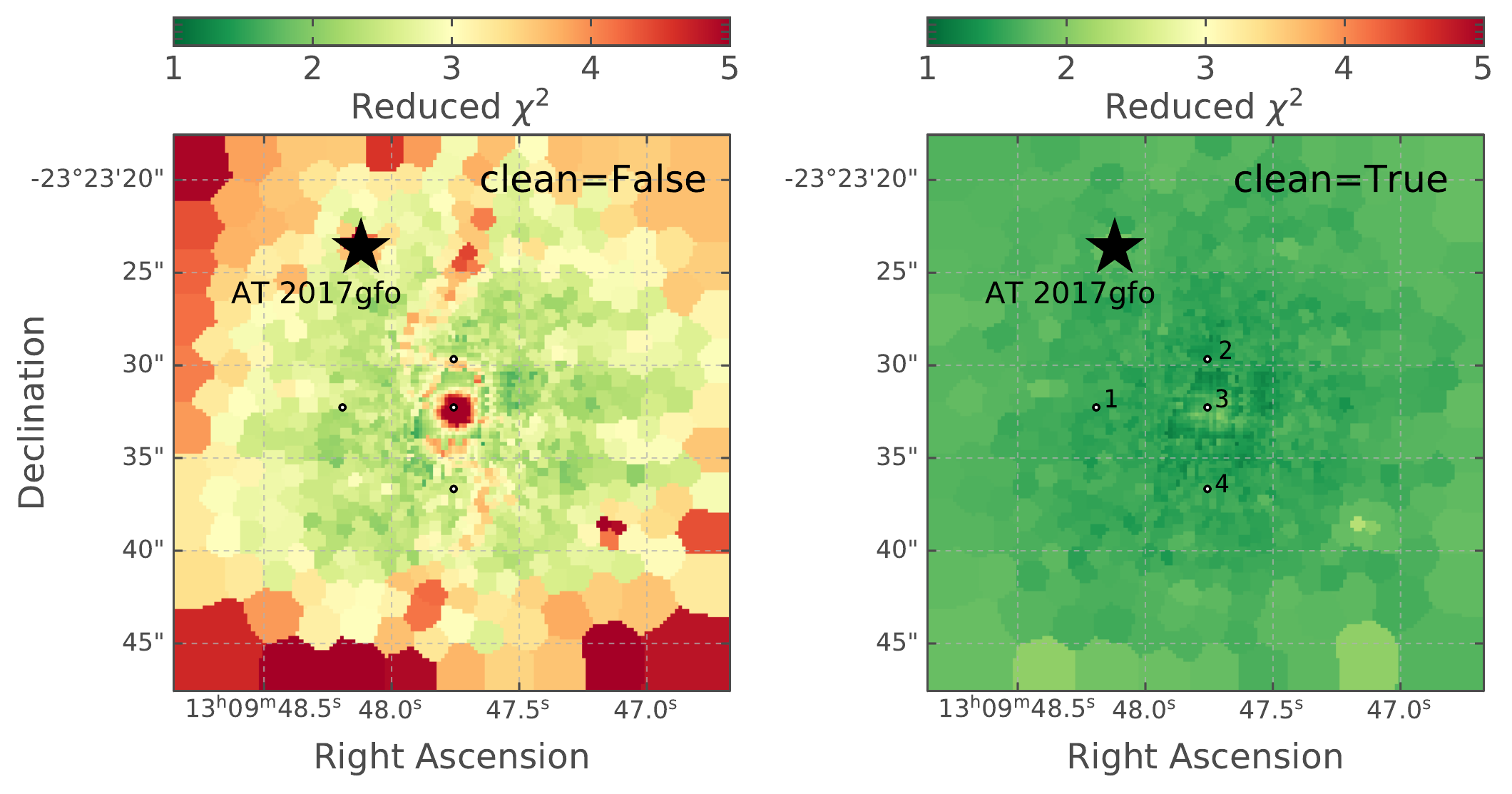}
    \includegraphics[width=17cm]{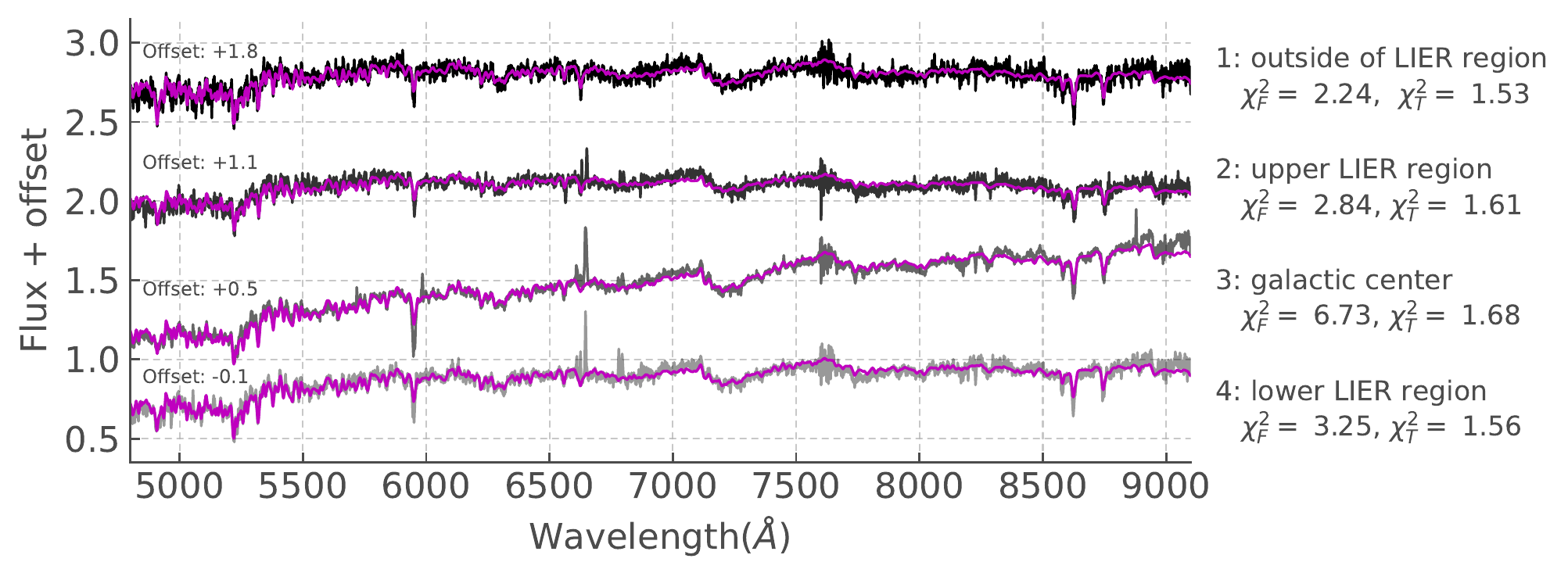}
    \caption{Comparison of the $\chi^2$ obtained when fitting with the clean=True or clean=False option. \textbf{Top Panels:} Maps of the reduced $\chi^2$ values obtained for fits performed with templates from 3 metallicities (Z=0.010, 0.020, 0.030), with (right) and without (left) sigma-clipping. \textbf{Bottom Panel:} Spectra from the voronoi bins at 4 different locations in NGC4993 (greys scale) and their fitted SEDs (magenta). Note that the shape of the fits with and without sigma-clipping is visually identical hence only one fit is seen. The 4 regions were selected to show representative spectra of NGC4993: (1) outside the region associated with ionised gas (LIER); (2) within the LIER region, at a location associated with high a $\chi^2$ value in the fit without sigma-clipping; (3) at the center of the galaxy where reddening and SNR are highest; (4)  within the LIER region, at a location associated with lower $\chi^2$ value in the fit without sigma-clipping (note that the emission lines are much weaker than in spectrum 2, resulting in the lower $\chi^2$). Finally, $\chi^2_F$ and  $\chi^2_T$ refer to the  $\chi^2$ values when the clean option is False and True, respectively.}
    \label{fig:chi2_clean_comp}
\end{figure}

\begin{figure}
    \centering
    \includegraphics[width=11cm]{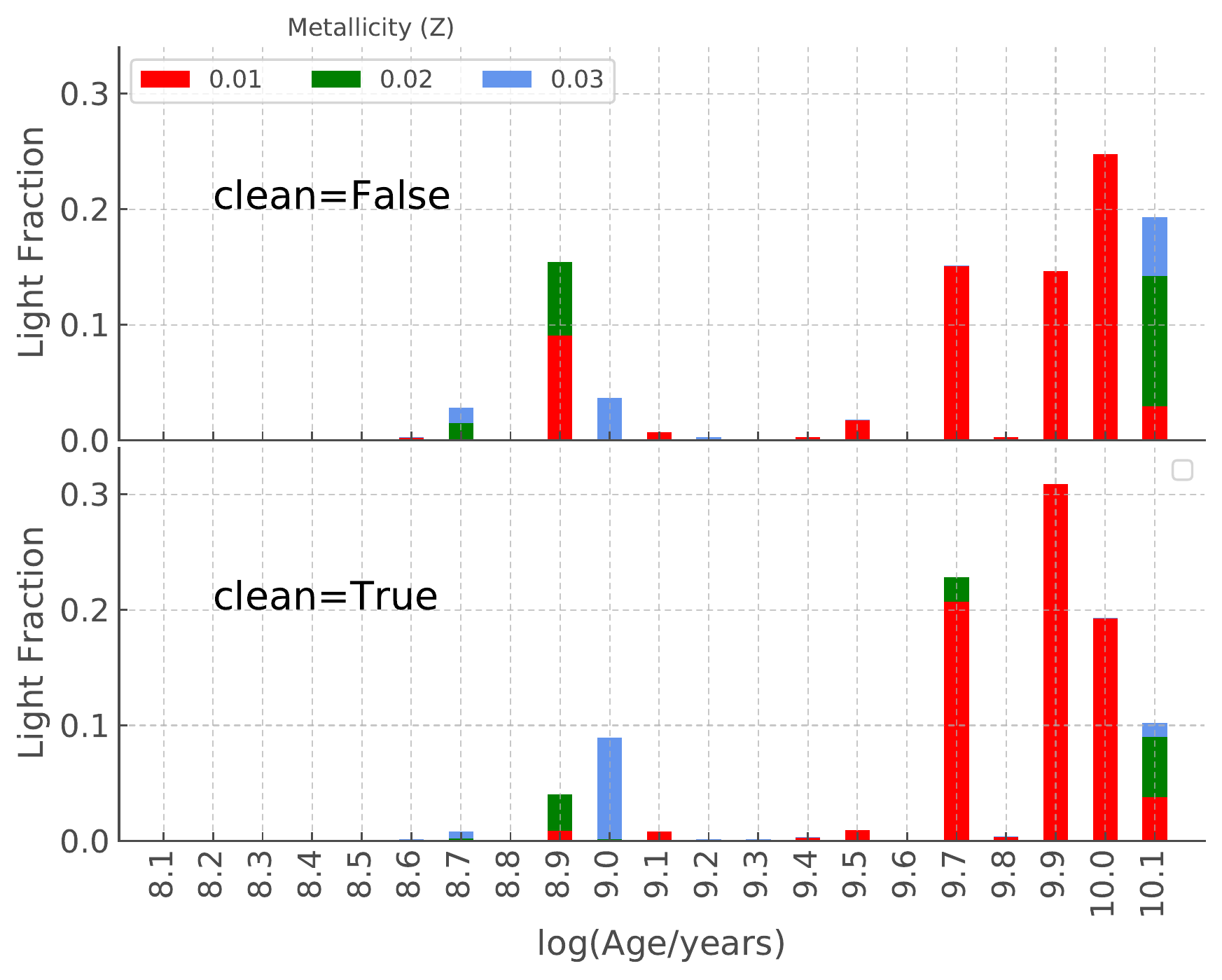}
    \caption{Star formation history of NGC4993. The fits were performed with templates from 3 metallicities with (bottom panel) and without (top panel) sigma-clipping.}
    \label{fig:sfh_clean_comp}
\end{figure}

The \ppxf\, SED fitting algorithm takes a number of input parameters, some required, some optional --for further detail on the \ppxf\, parameters see \cite{cappellari2004, cappellari2017} -- here we list the values of key parameters for our fits of NGC4993 presented in the Method section:
\begin{itemize}
    \item {\tt start} = [z$\times$c, 160\kms]: The two values of the {\tt start} parameters correspond to initial guesses for the recession and dispersion velocity. The former is taken as the overall recession velocity of NGC4993 (where z=0.009783 for NGC4993\cite{hjorth2017}) and the latter is an educated guess based on previous studies of this galaxy (and we find that \ppxf\,is consistent at retrieving the kinematic parameters even with start values differing by a few tens of \kms). 
    
    \item {\tt degree} = 2: This corresponds to the order of an \textit{additive} polynomial contributing to the continuum flux to be simultaneously fitted. Generally speaking a polynomial component can account for unconstrained issues in the flux calibration as well as reddening due to interstellar dust (to first order). 

    \item {\tt clean} = True: This refers to whether \ppxf\,is asked to perform sigma clipping on the spectral the data. We find extremely similar fits whether this option is enabled or not  -- the two major differences are in the reduced $\chi^2$ values obtained (see  Sup. Fig. \ref{fig:chi2_clean_comp}) and in the distribution of metallicities in the Star Formation History (SFH -- see  Sup. Fig. \ref{fig:sfh_clean_comp}) as e.g. narrow emission lines not modelled by the stellar model do not contribute to the $\chi^2$ calculations.
\end{itemize}

The $\chi^2$ on the top left panel of Sup. Fig. \ref{fig:chi2_clean_comp}, obtained through a fit that did not perform sigma clipping shows very interesting features. 
Firstly AT~2017gfo and a bright foreground star are clearly visible as clusters of high $\chi^2$ voronoi bins. 
Secondly, the "S"-shaped pattern across NG4993 corresponds to the spatial location of ionized gas associated with the LIER region previously reported: It is neither caused by the presence of an active galactic nucleus nor by young star formation regions \cite{levan2017}. 
Since we do not account for emission lines from ionised gas in our fit, the spectral features associated with the LIER region contribute to increasing the $\chi^2$ value -- unless sigma-clipping is performed in which case they are ignored. 
That is why this pattern is not observed in the reduced $\chi^2$ map from the fit performed with {\tt clean} = True. 
It is important to point out that in general it is not appropriate to ignore emission lines when performing SED fitting as they can be indicators of star formation regions and young populations (although they can also arise from interacting binaries in old stellar populations\cite{xiao2018}). 
In the particular case of NGC4993, independent evidence has already ruled out very young populations and identified those emission features as uncorrelated with star formation regions. 
This allows us set those features aside and keep the parameter space of our SED fitting as constrained as possible to focus on retrieving the Star Formation History (SFH).

Now, we bring our attention to the high $\chi^2$ region in the center of the galaxy. 
There are two effects at play: firstly the reddening in this part of the galaxy is stronger and more complex than we were able to fit, leading to a slight under-fitting of the flux for wavelengths greater than roughly 9000 \AA\, (see lower panel of Sup. Fig. \ref{fig:chi2_clean_comp}); secondly, the higher SNR (and smaller error-bars) in this region necessarily leads to greater $\chi^2$ - in fact the SNR pattern seen in Sup. Fig. \ref{fig:snr_voronoi} is discernible in both top panels of Sup. Fig. \ref{fig:chi2_clean_comp}. 
It is worth noting we also performed tests where reddening was simultaneously fitted, but this had no impact on the retrieved SFH in this instance. 

\begin{figure}
    \centering
    \includegraphics[width=17cm]{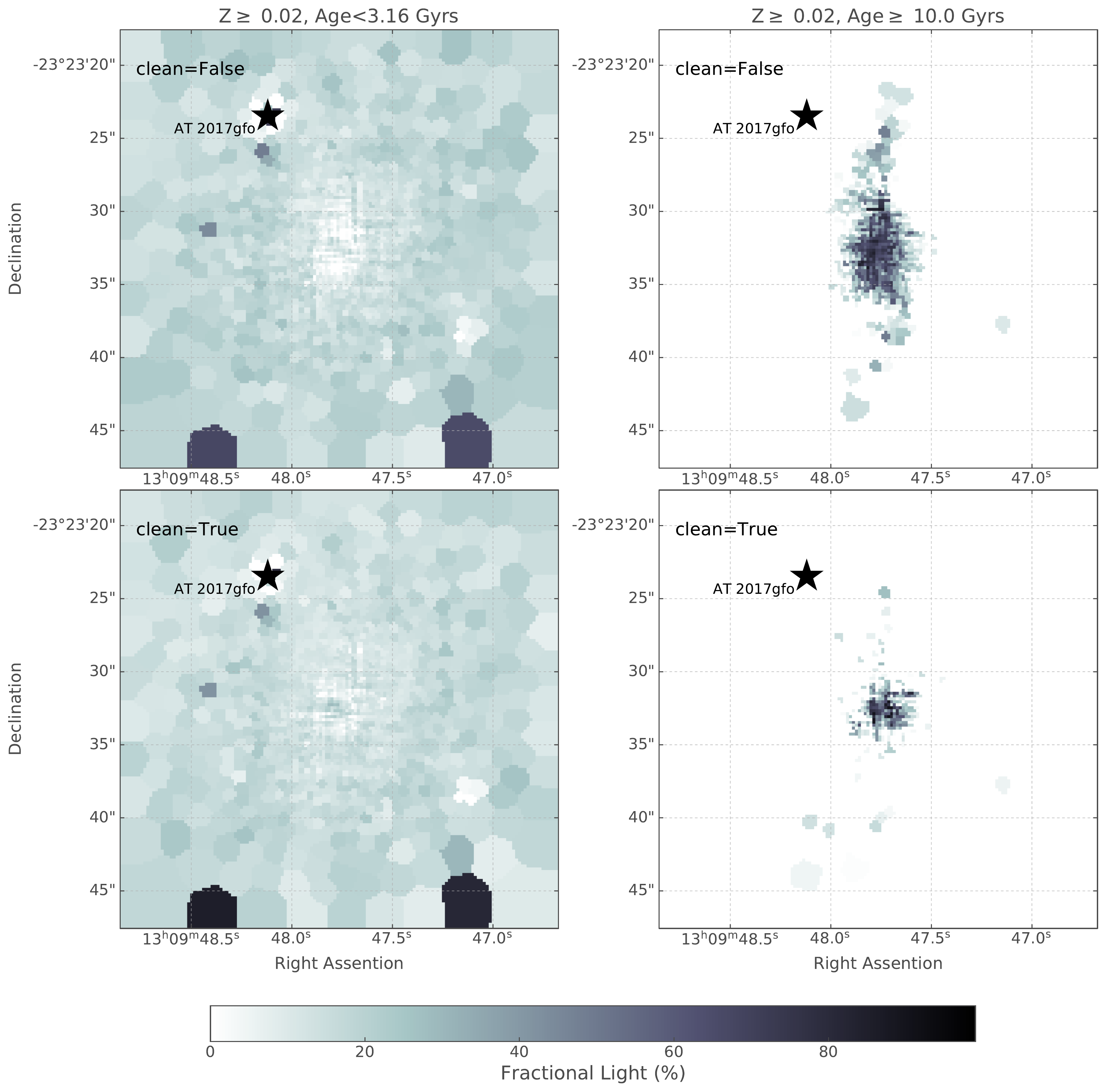}
    \caption{NGC4993 maps of the light fraction associated with the solar and super-solar templates. Left panels correspond to the templates associated with the younger population log(age/years)<9.5 (3.16 Gyrs) -- Right panels correspond to the templates associated with the oldest components log(age/years)$\ge$10.0 (10 Gyrs). The bottom and top panels are the results from the fits with and without sigma-clipping, respectively.}
    \label{fig:Z_clean_comp}
\end{figure}

The key difference between the {\tt clean} = True and {\tt clean} = False fits is one that could have an implication for the delay time search criteria we employ.
Sup. Fig. \ref{fig:sfh_clean_comp} reveals two interesting properties: a) the bimodal nature of the age distribution (with a population older than 3 Gyrs and a younger population with age around 1 Gyrs) is present in both fits; b) the number of high metallicity (Z=0.020 or Z=0.030) templates used in the 12.5 Gyrs bin is much greater in the fit without sigma-clipping.
The fact that metal rich populations would dominate star formation over the metal-poor (Z=0.010) population at the earliest stages of the Universe is counter-intuitive at best, so we investigate these components further.
In Sup. Fig. \ref{fig:Z_clean_comp} we show the light fraction of the Z={0.020, 0.030} templates with age greater than 10 Gyrs compared to the light fraction of the same Z templates at all other ages. 
We see that the suspicious high metallicity old population is concentrated in the areas of higher $\chi^2$ in both the fits with and without sigma-clipping.
On the contrary, the high metallicity templates with ages $<$3.16 Gyrs (i.e. those associated with the young population) are uniformly distributed throughout NGC4993 (apart from bogus peaks near AT2017gfo and in a couple of low SNR voronoi bins on the edges of the image).

Overall, we conclude that the solar and super-solar metallicity populations at ages > 10Gyrs are a numerical artefact.

%%%%%%%%%%%%%%%%%%%%%%%%%%%%
% constraining metallicities
%%%%%%%%%%%%%%%%%%%%%%%%%%%%%
\subsection{Constraining the metallicities used in our templates}
The templates we created to fit the SED of NGC4993 contained three (Z=0.010, 0.020, 0.030) of the 13 metallicities available in BPASS. 
Although in theory all ages and metallicity can be fit simultaneously, this can result in best fits with unphysical interpretation, such as an old population with a super-solar metallicity combined with a very young component with very low metallicity. 
This is because the recovery of star formation histories from SED fitting is an ill-conditioned inverse problem, meaning that it suffers from severe (and unsolvable) degeneracies \cite{cappellari2017}.

\begin{figure}
    \centering
    \includegraphics[width=11cm]{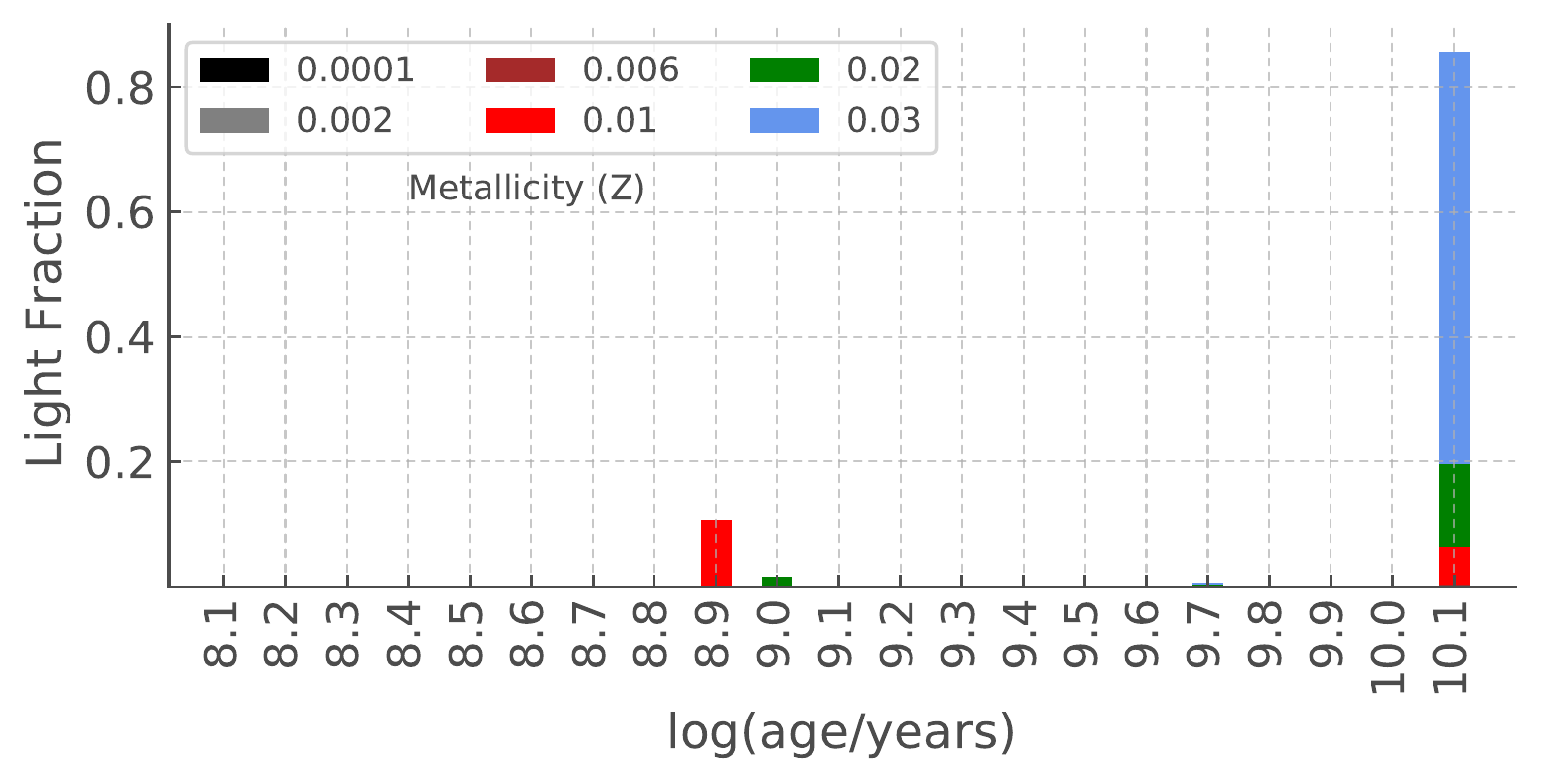}
    \caption{Star formation history of NGC4993 according to a fit performed with templates from 6 metallicities. Note that not all metallicities appear in the figure as they are not identified by \ppxf as needed to fit the data. }
    \label{fig:6_mets_sfh}
\end{figure}

To illustrate this we show the SFH resulting from one such fit, which was performed with templates for 6 metallicities (see Sup. Fig. \ref{fig:6_mets_sfh}). 
Although only 3 metallicities are clearly visible in the SFH, templates of all metallicities are present in the fit, but with weights so small their light fraction is not visible. 
According to this fit, NGC4993 would be dominated by a metal rich population created over 12.5 Billion years ago, with a large fraction of the metal poor (Z=0.010) population being less than 1 Gyrs old. 
This result is not physically sensible - it demonstrates that this type of SED fitting can be capricious when given too wide a parameter space and it is wise to constrain it using other observational clues or independent estimates. 

In the case of NGC4993, several metallicity estimates are available in the literature, which we summarise in Table \ref{tab:z_literature}. 
On the whole, NGC4993 is found to be slightly sub-solar to slightly super-solar, which is in accord with the metallicities dominating the 6 metallicity fit. 
Consequently we constrained our templates to Z=0.010, 0.020 and 0.030 to find our final SFH estimate of NGC4993.
Note that BPASS has an additional metallicity within this range (Z=0.014), and fits including these templates result in a very similar SFH with minimal contribution of the Z=0.014 templates. 
Since in BPASS solar metallicity is Z=0.020 and already included in our templates, the added degree of freedom brought by the Z=0.014 SEDs does not provide meaningful information to our analysis, hence we focused on the results obtained with the Z=\{0.010, 0.020, 0.030\} fits.

\begin{table}[h!]
    \centering
    \caption{Summary of previous estimates for the metallicity of NGC4993.}
    \label{tab:z_literature}
    \begin{tabular}{p{0.35\textwidth}p{0.60\textwidth}}

    %\multicolumn{2}{l|}{\textbf{Search criteria}} & \multicolumn{4}{l}{\textbf{Results}}  \\
    \textbf{Metallicity}  & \textbf{Method}  \\
    \hline
    \hline
    [Fe/H]=0.08, [Mg/Fe]=0.2 (1.2 - 1.6 \zsol) & Fitting of the spectrum in the nucelus of NGC4993 with a two-component SFH using MCMC \cite{blanchard2017}\\
    \hline
    [M/H]=-0.03 (\about0.9 \zsol) & Fitting of the integrated spectrum of NGC4993 with \ppxf and {\sc GANDALF}\cite{pan2017}\\
    \hline
    Z=0.02 and 0.05 (\about 1 - 2.5\zsol) & Fitting of the spectrum around AT2017gfo (2kpc from the center of NGC4993) with STARLIGHT models\cite{levan2017}\\
    \hline
   20\% - 100\% \zsol (best fit at 100\%) & Broadband SED Fitting \cite{im2017}\\
\end{tabular}
\end{table}

%%%%%%%%%%%%%%%%%%%%%
% COMPARISON TO JOE'S
%%%%%%%%%%%%%%%%%%%%%%%%%%%%%%%%%%%%%%%%%%
\subsection{Comparison to SFH previously inferred with this data-set}
\begin{figure}
    \centering
    \includegraphics[width=18cm]{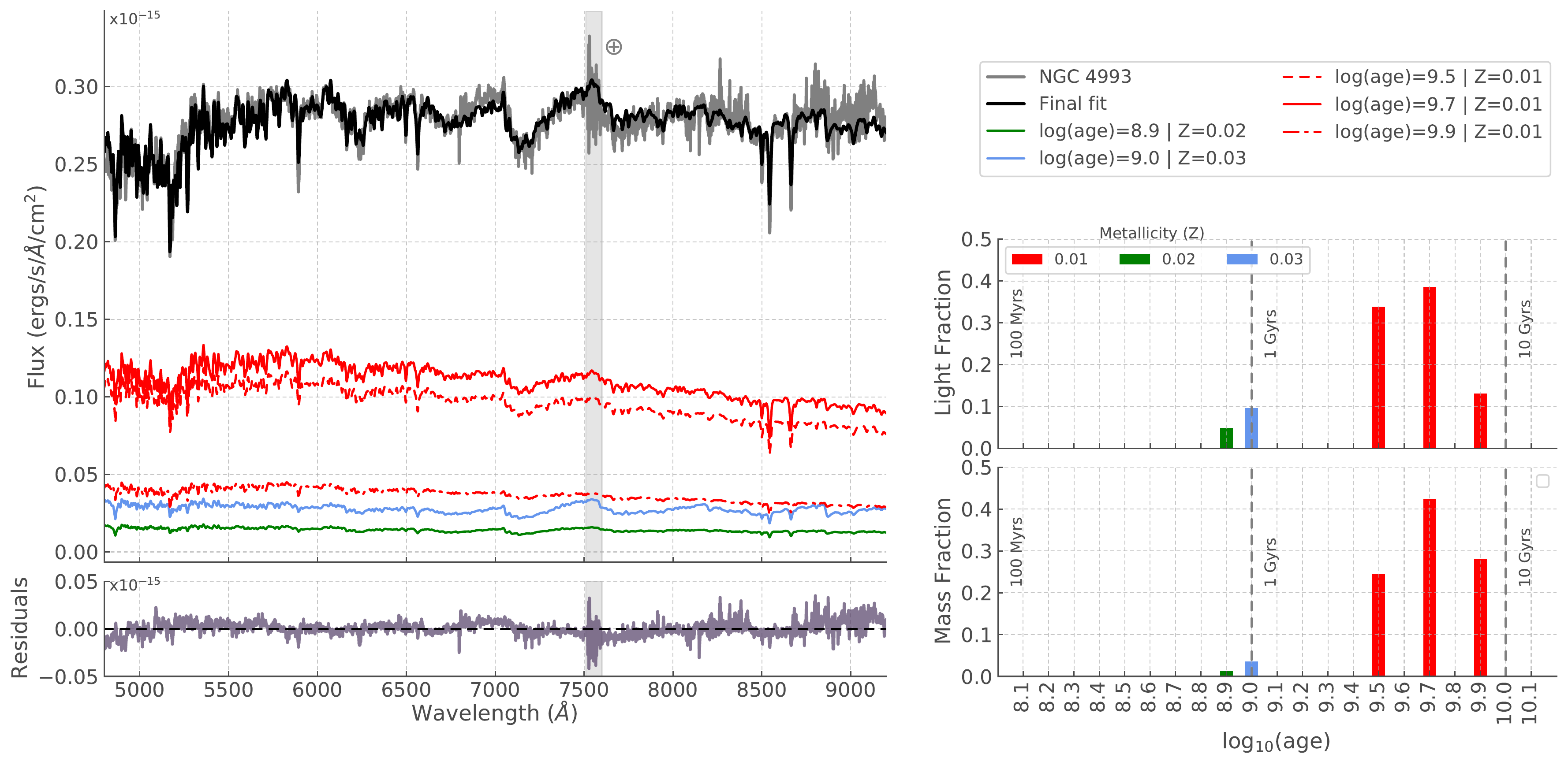}
    \caption{Best fit to the SED of the population in an annulus around the location of AT2017gfo and corresponding age and metallicity distribution. The coloured SEDs in the top left panel are the individual spectral components summed to obtain the final fit - the log age and metallicity are provided in the legend. The grey shaded area highlights a spectral region contaminated by telluric features that cannot be easily removed and the absorption line at 6795\AA\, is also a telluric remnant. The residuals have the same units as the flux. }
    \label{fig:sed_at_17gfo}
\end{figure}

Here we present a direct comparison to the SED fitting SFH previously reported for this data-set\cite{levan2017}, which focused on the region around the transient AT2017gfo and used SED templates that only considered single star populations. 
Here we do not use the voronoi tessalation; instead we integrate the flux of each pixel contained between a 1.5'' and a 2'' annulus centered on AT~2017gfo and the spectrum of the transient (integrated within a 0.5'' annulus) is removed to isolate the stellar component.
This follows the method described by the Levan et al. study\cite{levan2017}.
The templates metallicities and \ppxf settings are the same as those for the fits of the whole galaxy -- the resulting SED fit and derived SFH are shown in Sup. Fig. \ref{fig:sed_at_17gfo}. 
%In Table \ref{tab:sed@17gfo} we provide the values shown graphically in Sup. Fig. \ref{fig:17gfo}.

The age distribution found is very similar to that of NGC4993, and it also echos what was previously reported (see their figure 2\cite{levan2017}).
Another similarity is that the older population is found the be less enriched than the younger component, although in the original study the 1Gyr population has Z=0.050 and the \about 10 Gyr population has solar metallicity (Z=0.02). 
Additionally, they found the younger population to be more prominent than we did (over 20\% by mass in the bottom right panel of their figure 2 compared to just over 5\% mass fraction in the bottom right panel of Sup. Fig. \ref{fig:sed_at_17gfo}).

\begin{figure}
    \centering
    \includegraphics[width=13cm]{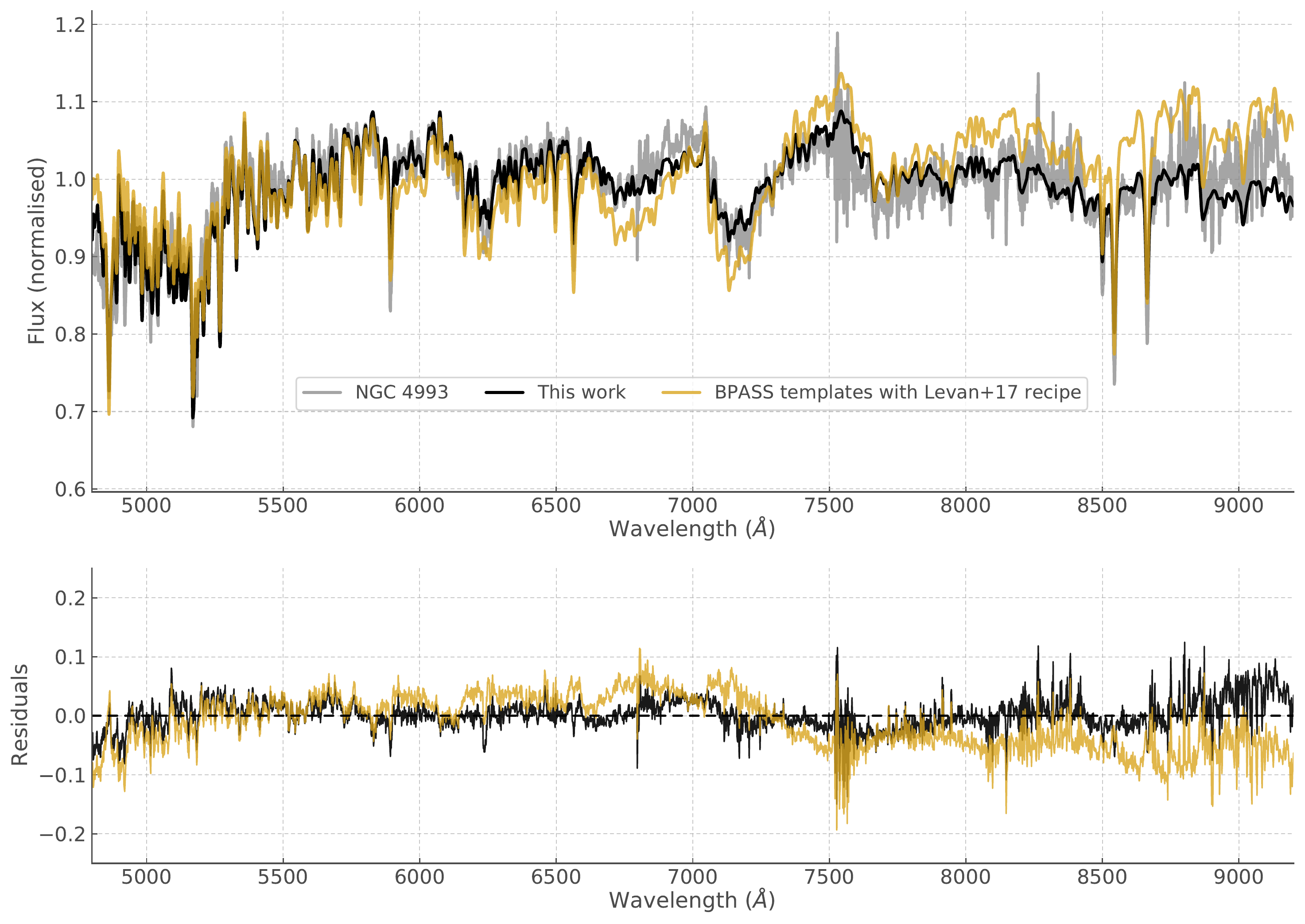}
    \caption{Best fit to the region around AT2017gfo using BPASS/hoki templates according to the SFH resulting from our SED fit (black) and that of Levan et al.\cite{levan2017}(gold). The larger metallicity leads to a strong TiO band, in fact our reproduction is an underestimate because we do not have Z=0.05 spectra and had to use a lower metallicity instead (Z=0.04). The bottom panel shows the residuals of both fits. The residuals are plotted in the normalised flux units.}
    \label{fig:me_vs_joe.png}
\end{figure}

To further investigate the differences between our two studies we create an analogue spectrum of their best fit using our templates. 
We take the SFH presented in their figure 2 and apply it to templates with Z=0.02 and 0.04 (BPASS does not include Z=0.05) - see Sup. Fig. \ref{fig:me_vs_joe.png}. 
Overall our Levan+17 analogue spectrum has much stronger TiO bands between 6600 and 7700 \AA\, and the flux in the near-infrared region is overestimated. 
We suspect the deviation in the fit is mainly due to the difference in the strength of the TiO bands from AGB stars in our BPASS models compared to the BC03\cite{bc03} stellar models that the original fit is based on\cite{levan2017, lyman2017}.
The modern stellar atmosphere models implemented in recent versions of BPASS have been previously described extensively - this includes an extensive discussion of the differences between BPASS and the BC03 models, which we refer the reader to\cite{stanway2018}.

\newpage

%%%%%%%%%%%%%%%%%%%%%%%%%%%%%%%%%%%%%%%%%%
%%%%%%%%%%%%%%%%%%%%%%%%%%%%%%%%%%%%%%%%%%
% BPASS   BPASS   BPASS   BPASS   BPASS
%%%%%%%%%%%%%%%%%%%%%%%%%%%%%%%%%%%%%%%%%%
%%%%%%%%%%%%%%%%%%%%%%%%%%%%%%%%%%%%%%%%%%
\section{Binary Population And Spectral Synthesis (BPASS)}
\label{sec:BPASS}
In the next subsection we describe the auxiliary numerical simulations that quantify the effect of kicks; we then consider the effects of the conversion from baryonic mass to gravitational mass in \ref{sec:NS_masses}, and detail how our models are weighted in \ref{sec:WEIGHTS}. 
Finally we provide an extensive discussion of common-envelope evolution (CEE) in BPASS in \ref{sec:CEE}.

%%%%%%%%%%%%%%%%%%%%%%%%%%%%%%%%%%%%%%%%%%
%%  TUI TUI TUI TUI TUI TUI TUI TUI TUI 
%%%%%%%%%%%%%%%%%%%%%%%%%%%%%%%%%%%%%%%%%%
\subsection{TUI: Supernova kicks and inspiral calculation}
\label{sec:tui}
The BPASS models in the standard library contain separate evolutionary tables for the primary and secondary stars, which we routinely call \textit{primary models} and \textit{secondary models}\cite{ghodla2022}.
Each evolutionary table has a unique combination of initial masses, period and metallicity.
In the case of the primary models, these quantities are the ZAMS values; in the case of the secondary models they correspond to the values of $M1, M2, P$ after the death of the primary star. 
In this context, we define a "genealogy" as a unique combination of primary model and secondary model.

This framework has advantages as well as disadvantages. 
The main drawback of splitting the primary star and secondary star evolution is that we cannot model binary systems with mass ratio equal to (or very close to) 1, as they will become evolved stars at similar times. 
This is not a major concern in general as binary systems born with equal masses are rare, but could be very relevant to BNS system formation (see discussion in \ref{sec:1CEE}).
On the other hand, our framework provides the advantage that it is trivial and computationally cheap to sample supernovae (SN) kicks and quantify their effect on our evolutionary channels. 
SN kicks are stochastic in nature and they can significantly change the orbit (and thus the period $P$) of a system.
As a result many primary models can lead to a single secondary model, and a single primary model can be followed by a number of possible secondary models, and the likelihood of these routes varies based on the kick distribution as well as the weighting inherent to the primary model from population synthesis.

TUI is the BPASS auxiliary code which performs Monte Carlo simulations of SN kicks and records which genealogies lead to the successful formation of compact binaries and subsequent mergers. 
It proceeds as follows: 
For each primary model, a random kick is drawn after the first SN and the new orbital parameters are calculated; one core assumption is that the orbit of the secondary models is ciruclar and so we circularise the orbits after the first SN kick using the semi-latus rectum as prescribed in the literature \cite{hurley2002}.
The new period is compared to the BPASS grid of periods of the secondary models (P2) and the closest match is used. 
In this study we randomly draw our kicks from a Maxwellian distribution with $\sigma=265$ \kms\cite{hobbs2005} for both SNe (see discussion in \ref{sec:kicks}), and 2000 random kicks are performed for \textit{each} SN.
We record how many times each secondary model is selected and normalise this value by the number of samples drawn from the kick distribution to give a fraction between 0 and 1, which we call $f_{P2}$.
After the second SN, a random kick is drawn again to calculate the orbital parameters of the new born BNS.
We record how many systems remain bound and normalise by the number of samples drawn to get the fraction of bound systems  $f_{\rm bound}$;
finally the orbital parameters are used to calculate the inspiral time using a recently derived approximation to the classic integration method in order to optimise computing time\cite{mandel2021a}.
This latter feature is a new addition of TUI.v3 which is used in this study (previously published works used TUI.v2\cite{ghodla2022}) 

%%%%%%%%%%%%%%%%%%%%%%%%%%%%%%%%%%%%%%%%%%
% MATCHING MODEL MASSES TO OBSERVED MASSES
%%%%%%%%%%%%%%%%%%%%%%%%%%%%%%%%%%%%%%%%%%
\subsection{Matching the modelled and observed masses of neutron stars}
\label{sec:NS_masses}
%%%%%%%%%%%%%%%%%%%%%%
\subsubsection{Gravitational and baryonic masses}
One of the key aspects of matching the observations of GW sources to the BPASS predictions is to compare the observed and predicted compact object masses. 
An important consideration is that the neutron star masses given by stellar models are \textit{baryonic} masses (\mb), i.e. the total mass of the matter comprising the neutron star. 
The observed mass, on the other hand, is the \textit{gravitational} mass (\mg), which is lower as it accounts for the binding energy of the compact remnant. 
The conversion from gravitational mass to baryonic mass is dependent on the Equation-of-State (EoS) adopted for neutron stars, and a commonly used approximation is:
\begin{equation}\label{eq:mbmg}
    M_b = M_g + A \times M_g^2
\end{equation}
where A has been calibrated for a number of Equation-of-State (EoS) and there exists an EoS-independent formalism\cite{timmes96} but the residual errors have been shown to be as large as 0.09\msol\cite{gao2019}. 
Additionally, these residuals are systematic errors which will differ in amplitude and behaviour across the NS mass range depending on the "true" EoS (see figure 1 of Gao et al. \cite{gao2019}). 
Consequently it would be misleading (and incorrect) to use these residuals to attempt to derive error-bars as is can be done with stochastic sources of uncertainty. 
In order to account for the effects of EoS choice we perform the conversion form baryonic to gravitational mass using 3 EoS (WFF1\cite{WFF1}, APR4\cite{AP4} and MPA1\cite{MPA1}).
They were chosen based on their overlap with the marginalised posterior distributions for the mass m and areal radius R calculated by the LIGO/Virgo collaboration\cite{theligoscientificcollaborationandthevirgocollaboration2018}. 
They show results for an EoS-indepent and with a parametrised EoS with a set lower limit on the maximum NS mass of 1.97 \msol: WFF1 covers the high density regions for the EoS-independent marginal posterior, whilst AP4 and MPA1 cover the two high density peaks of the posterior calculated with a limit on the maximum NS mass (see figure 3 \cite{theligoscientificcollaborationandthevirgocollaboration2018}).

We use the values of A reported in table 1 of\cite{gao2019} (0.102, 0.09 and 0.082 for WFF1, AP4 and MPA1, respectively) to solve Eq. \ref{eq:mbmg} and retrieve gravitational masses from our baryonic NS masses. 

%%%%%%%%%%%%%%%%%%%%%%555
\subsubsection{Weights from LIGO/Virgo mass posterior distributions}
\label{sec:LIGOweights}
\begin{figure}[h!]
    \centering
    \includegraphics[width=15cm]{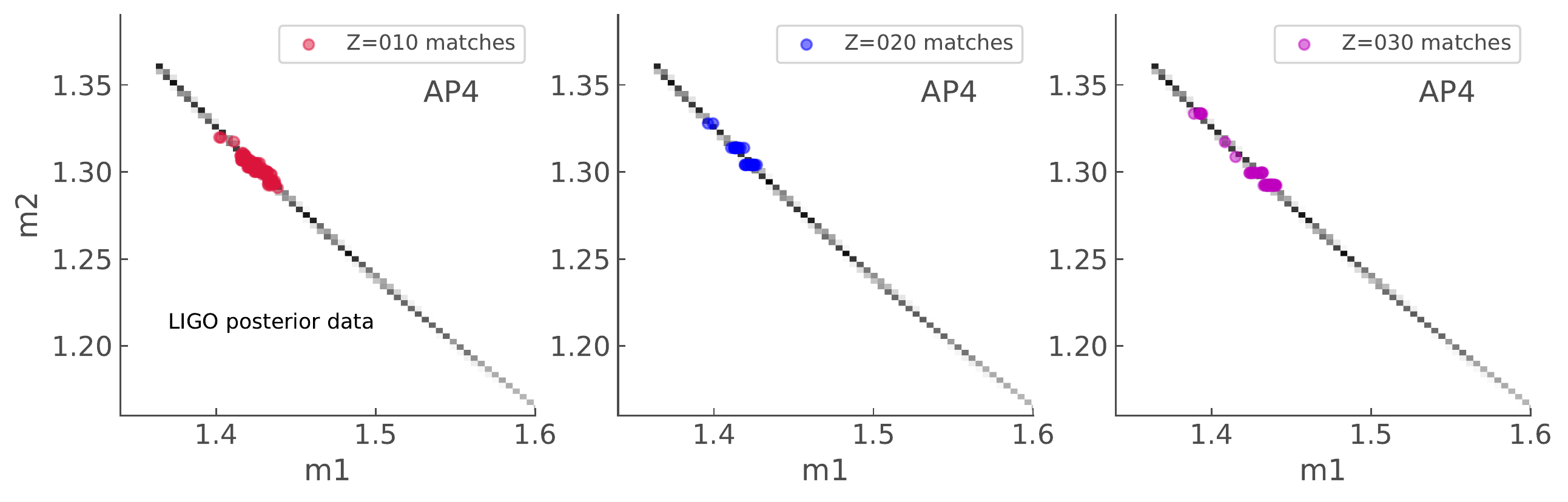}
    \caption{ $m_1$, $m_2$ space showing LIGO posterior distribution (greyscale) for the low spin prior and matching BPASS models (scatter plots) with baryonic mass correction performed with the AP4 EoS. Note that the BPASS models corresponding to posterior densities with 0 weight are not plotted. }
    \label{fig:match_m1_m2.png}
\end{figure}

Now that we have gravitational masses for our NS components it is possible to directly compare them to the measurements of the LIGO/Virgo team. 
We used the publicly available posterior distributions for GW170817 (see Data Availability) and create a 2D histogram in $m_1-m_2$ space with 100 bins in each dimension (an example is shown in Figure  \ref{fig:match_m1_m2.png}), where $m_1$ is the more massive NS, which is not necessarily the primary star (in fact we find that in most systems that is not the case). 
Each bin has a corresponding density which we include in the probability of our progenitor channels. 
In order to ensure the match to the LIGO/Virgo posterior distributions has a homogeneous impact on our conclusion \textit{regardless of EoS}, we normalise the densities to create a weight $w(m_1,m_2)$ by dividing each density in a sample (for a given EoS) by the maximum density for that sample.
The LIGO/Virgo posterior distributions for $m_1-m_2$ will vary depending on the prior used, here we consider both the low spin and high spin prior. 
Another effect we must account for is the redshift of the masses as the predictions for $m_1$ and $m_2$ are in the rest frame and the posterior distributions give the masses in the detector frame. 
A scaling factor of 1/(1+z) is applied to the LIGO/Virgo posteriors were the redshift of NGC4993 is taken as z = 0.009783\cite{hjorth2017}.
It is worth noting that even with this correction the priors used to by LIGO/Virgo are uniform over redshifted component masses but the impact on the posterior samples would be negligible compared to other sources of uncertainty in the modeling (stellar and natal kicks).

%%%%%%%%%%%%%%%%%%%%%%%%%%%%%%%%%%%%%%%%%%
% WEIGHTS WEIGHTS WEIGHTS WEIGHTS WEIGHTS
%%%%%%%%%%%%%%%%%%%%%%%%%%%%%%%%%%%%%%%%%%
\subsection{Quantifying progenitor channels and genealogies}
\label{sec:WEIGHTS}
Earlier we defined a "genealogy" as a unique combination of primary and secondary model. 
It is important to note that we differentiate between genealogies and evolutionary channels - the former is quantitative and BPASS specific, the latter is (mostly) qualitative and applicable across stellar evolution. 
Two different genealogies may show the same channel, say, the occurrence of two CEE, but will differ in the age and timestep at which RLOF occurs and how much mass is lost. 
In order to quantify which evolutionary channels are preferred following our search criteria, we must first weigh each genealogy. 
To this end we define the following:
\begin{equation}
   w_{\rm gen} = N_{\rm}(M_{1,\rm zams}, M_{2,\rm zams}, P_{\rm zams})^{Z} \times f_{\rm P2}^{Z} \times f_{\rm BNS}^{Z} \times f_{\rm mass}^{Z} \times w(m_1, m_2), 
\end{equation}

where $N_{\rm}(M_{1,\rm zams}, M_{2,\rm zams}, P_{\rm zams})$ is the expected number of systems with ZAMS $M_1$,  $M_2$,  and period $P$ for a 10$^6$\msol population with metallicity Z; $f_{\rm P2}$ is the selection fraction of a secondary model (as defined in Section \ref{sec:tui});  $f_{\rm BNS}$ is the fraction of bound BNS resulting from a particular genealogy (see Section \ref{sec:tui}); $f_{mass}$ is the mass fraction of a population according to our SED fitting results (see Section \ref{sec:SEDfit})); and $w(m_1, m_2)$ is the weight calculated from the LIGO/Virgo posterior distributions (see Section \ref{sec:LIGOweights}). 
The Z upper script indicate the metallicity dependent quantities. 

Once $w_{\rm gen}$ have been calculated, they can be grouped, summed and normalised to retrieve the percentages we quote for the evolutionary channels in Section \ref{sec:results}

%%%%%%%%%%%%%%%%%%%%%%%%%%%%%%%%%%%%%%%%%%
%%  CEE CEE CEE CEE CEE CEE CEE CEE
%%%%%%%%%%%%%%%%%%%%%%%%%%%%%%%%%%%%%%%%%%
\subsection{The Common Envelope}
\label{sec:CEE}
\subsubsection{Brief summary of the parameterisation of common envelope in the literature}
Common envelope evolution (CEE) is one of the critical steps in the formation of BNS\cite{bhattacharya1991, tauris2006, belczynski2007, tauris2017, kruckow2016, kruckow2018, chruslinska2018, belczynski2018, belczynski2018a, vigna-gomez2018, vigna-gomez2020, olejak2021} and it is also the main source of uncertainty.
A CEE phase arises when mass transfer is dynamically unstable as a result of the donor star expanding (relative to the Roche Lobe) in response to the mass transfer, or when the accretion rate is so large relative to the accretor's reaction timescale that matter fills the binary orbit (e.g. neutron star fed at super-Eddington rates\cite{king1999}).
The filling of the binary orbit first results in a rapid plunge-in phase (which occurs entirely on the dynamical timescales).
The spiraling of the companions during this phase transfers orbital energy to the envelope, driving its expansion - if it is sufficient, then the system enters a self-regulating spiral-in phase, which operates on the thermal time-scale of the envelope, and ends either with a slow merger or with the ejection of the envelope. 
If the self-regulating phase cannot be entered the binary merges rapidly, unless it is able to eject its envelope on short timescales.

The key point we need in order to assess a given CEE phase's consequence on the life of our progenitors is \textit{"What is the resulting orbit shrinkage?"} as this will determine how many systems merge within a given delay time. 
But before this can be quantified, many other questions must be asked relating to the total energy budget (including sinks and sources), and energy and angular momentum transport within the envelope. 
(For a review of the field see \cite{2020cee..book.....I}).
The physics of CEE spans many orders of magnitudes in time and spatial dimensions,  so parameterisations have been developed to approximate this phase and make it accessible to current computational capabilities.
There are two common approaches:
\begin{itemize}
    \item The $\alpha$--formalism (energy conservation formalism\cite{webbink1984, dewi2000} ):
    \begin{equation}
        E_{\rm binding}=\alpha_{\rm CE} \left( E_{\rm orb,i} - E_{\rm orb,f} \right),
    \end{equation}
    where $E_{\rm binding}$ is the binding energy of the donor envelope and the other energy terms are the initial and final orbital energies. While the exact for of this expression may vary with the specific implementation it is generally written as:
    \begin{equation}\label{eq:alpha}
    \frac{GM_{1, \rm env} M_1}{\lambda R_1} = \alpha_{\rm CE} \Big(-\frac{G M_1 M_2}{2a_i} + \frac{G M_{1,c} M_2}{2a_i} \Big),
    \end{equation}
    where  $M_2$ is the mass of the accretor, $M_1$ is the mass of the donor star , $M_{1, \rm env}$ is the mass of its envelope,  $M_{1, \rm c}$ is the mass of its core (such that $M_1=M_{1, \rm env}+M_{1, \rm c}$); $a_i$ and $a_f$ are the binary separation at the onset and at the end of CEE, respectively; $\lambda$ accounts for how the structure of a star affects its binding energy (which will change across the evolutionary stages of a star\cite{dewi2000}), and $\alpha_{\rm CE}$ is the CEE efficiency which is meant the represent how much of the orbital energy is being transferred to the envelope to unbind it. 
    
    \item The $\gamma$--formalism (angular momentum conservation\cite{nelemans2000}):
    \begin{equation}\label{eq:gamma}
    \frac{\Delta J_{\rm lost}} {J_i} = \gamma \frac{M_{1,env}}{M_1 + M_2},
    \end{equation}
    where $J_i$ is the angular momentum, and $\gamma$ is the efficiency of angular momentum transport.  
\end{itemize}

Additional approximations of CEE in stellar population synthesis codes are often found in the determination of the onset of runaway mass transfer.
These conditions can be more or less complex and typically make use of mass ratio thresholds\cite{olejak2021}.
Another common generalisation is that all systems which enter CEE with an Hertszprung-Gap (HG) donor will merge \cite{belczynski2007, compas2021}.
To keep this discussion (relatively) succinct we focus on our BPASS implementation, but we encourage the interested readers to refer to the following key references for further detail on how other stellar evolution codes treat CEE:  {\tt StarTrack}\cite{belczynski2002,belczynski2008}, {\sc ComBine}\cite{kruckow2018}, COMPAS\cite{compas2021}, MESA\cite{paxton2011,paxton2015,paxton2018,paxton2019}. 

\subsubsection{The BPASS-STARS implementation}
The detailed stellar evolution modeling in BPASS is performed using a custom version of the Cambridge STARS code \cite{eggleton1971,eldridge2008}, and our treatment of CEE differs in several ways to other established stellar evolution codes.
One of the first points that sets us apart is in our prescription for the onset of CEE. 
We do not use mass transfer stability criteria based on e.g. the phase of evolution, mass ratio (q) thresholds, or donor mass/radius. 
Instead we allow mass transfer to occur once the donor star fills its Roche lobe. We only consider  that CEE has begun when the radius of the donor is greater or equal to the separation ($R_1\ge a$). Thus significant orbital evolution occurs before we begin our CEE prescription that would typically be included as CEE in other prescriptions. Detailed calculations of the stellar interior with STARS continue throughout the CEE phase: the mass-loss rate is allowed to increase (up to a value of 0.1 $M_{\odot}.{\rm yr}^{-1}$ for numerical stability\cite{eldridge2017}), and the envelope mass and orbital separation are updated. %\textcolor{gray}{ (see section 2.1.2 and equation 7 of \cite{eldridge2017}). Our formulation is similar in spirit to the $\alpha$--formalism as energy conservation between the binding and orbital energy ($E_{\rm bind} \approx \Delta E_{\rm orbit}$) is the basis for our analytical formulations, but the step-by-step calculations of the detailed models means that we do not have a set value of $\alpha_{CE}$ -- the efficiency will vary between models and for each star over the course of their CEE phase. }
The CEE implementation within the code assumes that at each time-step that $dJ \approx J\, dM_1/(M_1+M_2)$, similar to the $\gamma$-formalism. Previously we thought the implementation has been similar to the $\alpha$-formalism (see section 2.1.2 and equation 7 of \cite{eldridge2017}). We have confirmed that this is not the case within the stellar models and that it is angular momentum that is conserved. We note this is only the case during the CEE phase as defined within BPASS where it is likely to be at least comparable to conserving energy (section 9.4\cite{2020cee..book.....I}). We also note that because of the step-by-step calculations of the detailed models means that we do not have a set value of $\gamma_{CE}$ or effective $\alpha_{CE}$ -- thus the efficiency varies between models and for each star over the course of their CEE phase.

This step-by-step calculation is a crucial difference because many CEE prescriptions use quantities calculated at the onset of CEE to infer its final result, but if this phase occurs on the thermal timescales (i.e. if a self-regulating spiral-in takes place) then changes in the stellar interior structure cannot easily be ignored.
In principle this approach will provide better approximations of the effects of CEE as it allows the donor star to respond to the interaction and the binding energy to be updated.
One of the corollaries of the standard $\alpha$-formalism is that the kinetic energy is imparted to the envelope evenly and all of the envelope receives just the right amount of energy to leave the system at the escape velocity. 
In practice this will not be the case - even if the kinetic energy imparted on the envelope as a whole is exactly that required for all mass to leave the system at the escape velocity, the energy is not distributed homogeneously and the outer layers will leave with greater speed whilst the inner layers will fallback on the system (see the discussion on enthalpy\cite{ivanova2011, ivanova2013}). 

\begin{figure}[h!]
    \centering
    \includegraphics[width=11cm]{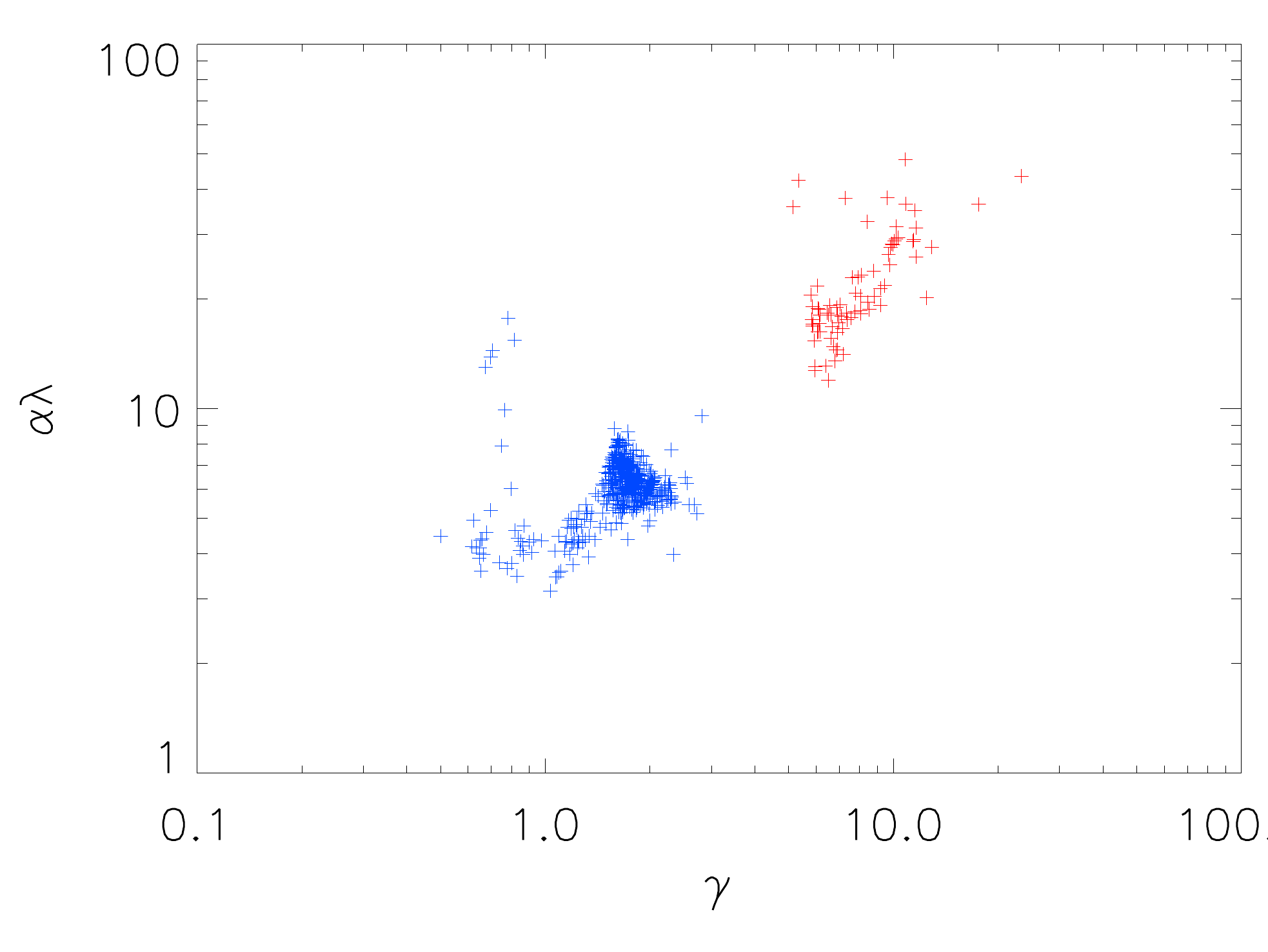}
    \caption{$\alpha_{CE}$ against $\gamma$ for the CEE of primary models (blue) and secondary models (red) that match our criteria, for all EoS and metallicities.}
    \label{fig:alpha_gamma}
\end{figure}

To allow for further discussion and comparison to the typical CEE prescriptions we calculate the effective $\alpha_{CE} \lambda$\footnote{Since we do not directly use the formalisms described in the previous section we cannot formally calculate separate values of  $\alpha_{CE}$ and $\lambda$.} and $\gamma$ of the CEE phases undergone by our matching models (see Section \ref{sec:results}), see Sup. Fig. \ref{fig:alpha_gamma}. 
Here we have calculated values of $\alpha_{CE} \lambda$ and $\gamma$ using the relevant stellar and orbit parameters from the onset of Roche-Lobe Overflow (RLOF), to when CEE ends. This is as direct a comparison as we can make to other CEE implementations. 
In previous comparisons we have only considered these values from calculations using the start and end of the CEE phase within our prescription. 
This did not take into account the significant evolution that occurs at the beginning of mass transfer before the spiral-in phase, such as the impact of the mass-transfer itself and for the primary models the transfer of angular momentum from the orbit to the stars by tides.

In the primary models (where the accretor is not a compact remnant) the $\gamma$ values span a range from ~0.5 to 3 which is rather consistent with values quoted in early studies \cite{2005MNRAS.356..753N}, whilst the $\alpha_{CE} \lambda$ values range from about 2 to 10 with a peak ~5, meaning that from an $\alpha$-formalism perspective the energy budget is consistently overspent. 
This need for $\alpha_{CE}\lambda>1$ has long been noted in the literature and can be a consequence of the fact that many energy sources, beyond the binding energy, are not taken into account in the $\alpha$-formalism, such as recombination and nuclear energy.
Additionally, the internal energy of a star can be critical -- although it is not an energy source \textit{per se}, it decreases the effective binding energy and therefore the amount of orbital energy required to eject the envelope.
3D hydro-dynamical simulations of 12 \msol super-giants have showed that including radiation energy results in the ejection of 60\% more mass, and including recombination energy as well results in final separations increased by 20\% \cite{lau2021}. 

Another key consideration is the uncertainty in the determination of the $\lambda$ parameter.
An important factor is the core-envelope boundary, also-called the bifurcation point: $\lambda$ is very sensitive to it \cite{tauris2001} and changes in the prescription of this bifurcation point can lead to large changes in the effective $\lambda$ value, especially for steep density gradients near the core\cite{kruckow2016}.
In BPASS the core-envelope boundary is defined as the location in the stellar interior where the hydrogen mass fraction $X_H \le$10\%. 
This core-envelope boundary criterion is commonly used \cite{dewi2000, kruckow2016} but it is also often taken to be equivalent to the bifurcation point.
This is not the case in BPASS. Our bifurcation point includes both the core mass and a small amount of envelope mass - a critical mass where the envelope would be unstable to being a giant if considered in isolation. 
As a result we find many examples of stars where after significant mass-loss from CEE, the left-over envelope shrinks back onto the star, bringing unstable mass transfer to an end and leaving behind a \textit{stripped, but not naked}, core.
This is another key difference between BPASS and codes which assume complete ejection of the envelope (e.g. COMPAS\cite{compas2021}).
Overall, variations in the stellar structure (and therefore binding energy) can also be responsible for our range of $\alpha_{CE}\lambda$ values -- in figure 5 of \cite{ivanova2013} the $\lambda$ value alone can change by an order of magnitude between scenario where the bifurcation point is X=0.1 and other prescriptions.%\cite{han1994}.

The take home points are as follows: CEE in BPASS quantitatively different from other prescriptions in the literature -- it follows the early phase of mass transfer rather than immediately jumping to CEE, it is more efficient, leads to less orbital shrinkage, and  allows a small portion of the envelope to shrink back onto the donor. 
BPASS is able to simultaneously reproduce a large array of observables, such as runaway star velocities and masses\cite{eldridge2011}, the relative rates of different SN types and ratios of different massive star types\cite{eldridge2017}, and the rates of CCSNe, type Ia SNe, GW transients\cite{eldridge2019,briel2022}. 
Although this brings us confidence in the BPASS predictions and general results, this cannot be taken to mean that our prescription is necessarily better than other prescriptions in the literature.
Many degenerate effects are at play and so the details may be incorrect whilst large scales results remain consistent with observational data. 
For example, we know and have previously reported that our CEE timescales are overestimated\cite{eldridge2017} -- this comes as a result of a mass-loss rate cap of 0.1 $M_{\odot}.{\rm yr}^{-1}$ (to avoid numerical issues).
Additionally, the secondary models, which undergo CEE with a NS, have much larger values of $\alpha_{CE}\lambda$ than the primary models - we could imagine that the hard radiation of the NS may provide an additional source of energy (of order the Eddington luminosity) to eject the envelope, but at this stage this is not formally included and will be quantified in future work. However, much of this larger $\alpha_{CE} \lambda$ is the result of tides not being taken into account in our secondary models. This is because of the assumption that the companion stars are already expected to be rotating at a similar period to the orbital period compared to when on the ZAMS. Thus the evolution of the orbit at the beginning of mass transfer is less than we find in our primary models.
%Additionally, STARS currently does not take into account stellar rotation, which could be highly relevant to CEE phases with a compact remnant which may have been previously processed by stable (or quasi-stable) mass transfer.
These uncertainties would result in different pre-supernova periods, but the exact impact on the progenitor population is not easily quantifiable, particularly given the interplay between orbit shrinkage and supernova kicks: a variation in CEE outcomes could simply lead to a different conclusion on what kicks magnitudes/orientation a system need to undergo in order to match a given observational criteria (in our case, merger time).

We will be investigating more options and variations on our STARS CEE formalism in an upcoming study to better quantify the variations in quantities described above, but a true understanding of CEE will require further work from many teams and across different methodologies (hydro-dynamical simulations as well as population synthesis) -- it will take many years to bring the field to consensus on how to best approximate this complex phase of a star's life. 
These unavoidable uncertainties in stellar evolution are another example of why we have developed the framework to perform end-to-end analysis of transient host populations and their progenitor: in trying to understand the genealogies of stars, approximations must be made, and here we can ensure our assumptions are homogeneous throughout.

%%%%%%%%%%%%%%%%%%%%%%%%%%%%%%%%%%%%%%%%%%
%%%%%%%%%%%%%%%%%%%%%%%%%%%%%%%%%%%%%%%%%%
% GENEALOGIES EXTENDED DISCUSSION
%%%%%%%%%%%%%%%%%%%%%%%%%%%%%%%%%%%%%%%%%%
%%%%%%%%%%%%%%%%%%%%%%%%%%%%%%%%%%%%%%%%%%
\section{Progenitor channels of GW170817}
\label{sec:results}

After matching our models to our selection criteria and applying the weighting described in Section \ref{sec:WEIGHTS} to each genealogy, we can now compare and quantify the favoured evolutionary channels that lead to candidate progenitors of GW170817. 
The values of $w_{\rm gen}$ are normalised by the sum of the weights over all three metallicites (Z=0.01, 0.02, 0.03) for one given EoS and prior pair (low spin or high spin prior). 
From this point on, it is implied that our weights are normalised weights. 
We find that for the Z=0.010 population totals 99.0\%,  98.8\%, and 98.8\%of the weights in the results found for WWF1, MPA1 and AP4, respectively, when using the LIGO/Virgo posterior masses calculated with a low spin prior. The high spin prior returns even higher weights toward the Z=0.010 for all three EoS.
In the following descriptions we therefore focus on the Z=0.010 solutions, as they are overwhelmingly likely to be the progenitors of GW170817. 
In a future study we will explore the metallicity dependence of our genealogies in the wider context of binary neutron star mergers, irrespective of whether they match GW170817 characteristics.

Unsurprisingly, the choice of EoS has an impact on the genealogies we retrieve as well as the number of potential progenitor systems. 
WFF1 returns the fewest models, with only 48 (95) unique systems, compared to 251 (391) and 119 (282) different systems for MPA1 and AP4 when using the low spin (high spin) prior. 
In fig3 of \cite{theligoscientificcollaborationandthevirgocollaboration2018} WFF1 just grazes the marginalised posterior distribution in m-R space when a lower limit on the maximum NS mass is imposed (1.97\msol - see their right panel), while MPA1 and AP4 pass through the densest regions. 
It is therefore possible that WWF1 is a less appropriate EoS, therefore returning fewer potential progenitors and with genealogies that are less in line with the other two EoS (see below).

In Sup. Fig. \ref{fig:flowchartMPA1}, \ref{fig:flowchartAP4} and \ref{fig:flowchartWFF1} we present diagrams summarising key features of the progenitor genealogies retrieved for EoS MPA1, AP4 and WFF1, respectively.
The evolutionary channels are summarised with abbreviations in the legends, where 1 refers to the primary and 2 refers to the secondary.
We indicate the CEE and stable mass transfer (STM) phases alongside with the general evolutionary phase during which they occur: cases A, B or C. 
In BPASS, these are defined programatically with the following thresholds:
\begin{itemize}
    \item \textbf{Case A}: Onset of RLOF before the He core mass coordinate reaches 0.1 \msol (during the main sequence).
    \item \textbf{Case B}: Onset of RLOF when the He core with mass coordinate > 0.1\msol and the CO core mass coordinate < 0.1\msol (after the formation of a helium core - which includes hydrogen shell burning).
    \item \textbf{Case C}: Onset of RLOF for stars which have CO core mass-coordinates > 0.1 \msol (after formation of a carbon-oxygen core, which includes helium shell burning and beyond). Some case C stable mass transfer (SMT) phases would be classified as case BB in the literature if the star has undergone RLOF previously - we will maintain a nomenclature that is independent of the previous RLOF phases that may have occurred, but the fact that case BB SMT is not mentioned in our channels is only a matter of convention.
\end{itemize}
So for example the channel ``1(CEEB) \& 2(CEEB+SMTC)" is a system where the primary star underwent case B common envelope evolution, died, then the secondary underwent case B common envelope and then another phase of stable mass transfer  late in life; \textit{the latter is sometimes called case BB mass transfer - here we do not use this notation to avoid switching from the ``case C" to ``case BB" and maintain consistency.
It is also worth noting that the colours in Sup. Fig. \ref{fig:flowchartMPA1}, \ref{fig:flowchartAP4} and \ref{fig:flowchartWFF1} have been coordinated to help visualisation: blue hues show systems that undergo CEE in the primary and the secondary, red hues show systems where the primary undergoes stable mass transfer on the main sequence and green hues show systems where the secondary undergoes two phases of common envelope. 
}
Note that the sum of the weights do not necessarily add to the population totals quoted above -- the remaining weights are found to be scattered across very low probability genealogies with large initial periods (>250-400 days).

\begin{figure}[h!]
    \centering
    \includegraphics[width=7cm]{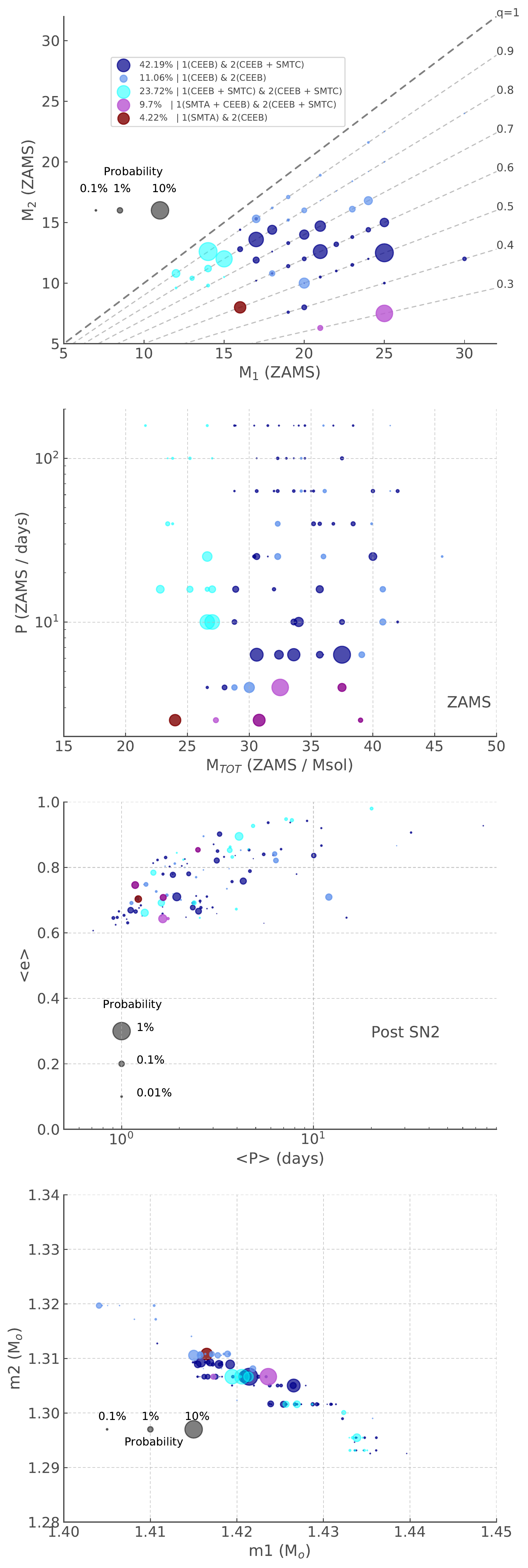}
    \includegraphics[width=7cm]{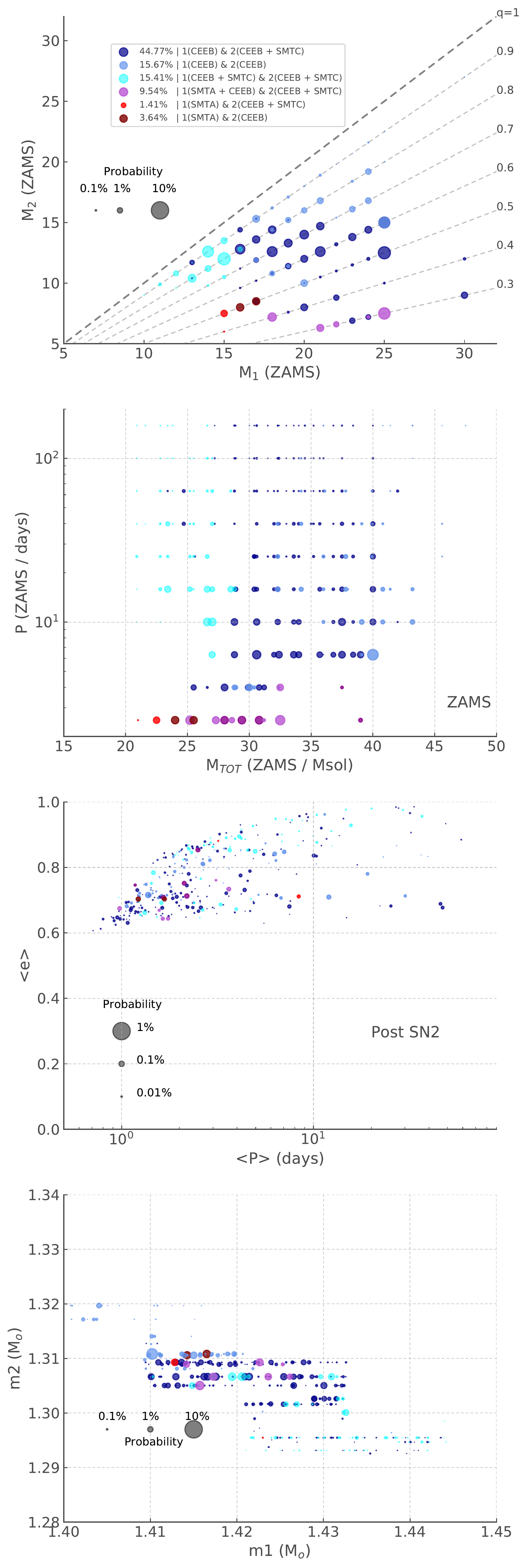}
    \caption{Summary proporties of the candidate progenitors of GW170817 in BPASS using the baryonic mass correction for the MPA1 EoS. \textbf{Left}: Results from comparing to $m_1$-$m_2$ calculated using the low spin prior. \textbf{Right}: Results from comparing to $m_1$-$m_2$ calculated using the high spin prior. Note that the size of the circles in the legend correspond to the average size plotted, and is not indicative of the total fraction of each channel, which is given in percentages instead. A comparatively large circle and a small percentage indicates that most of the probability is contained in only a few points rather than spread across the parameter space.}
    \label{fig:flowchartMPA1}
\end{figure}

\begin{figure}[h!]
    \centering
    \includegraphics[width=8cm]{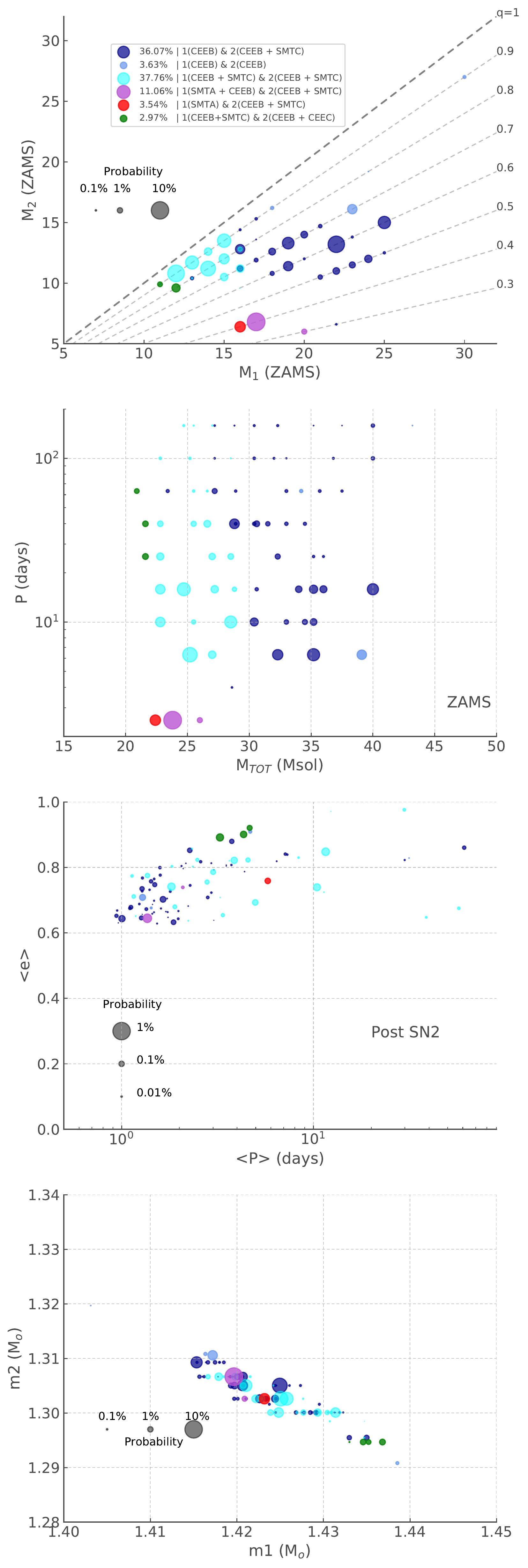}
    \includegraphics[width=8cm]{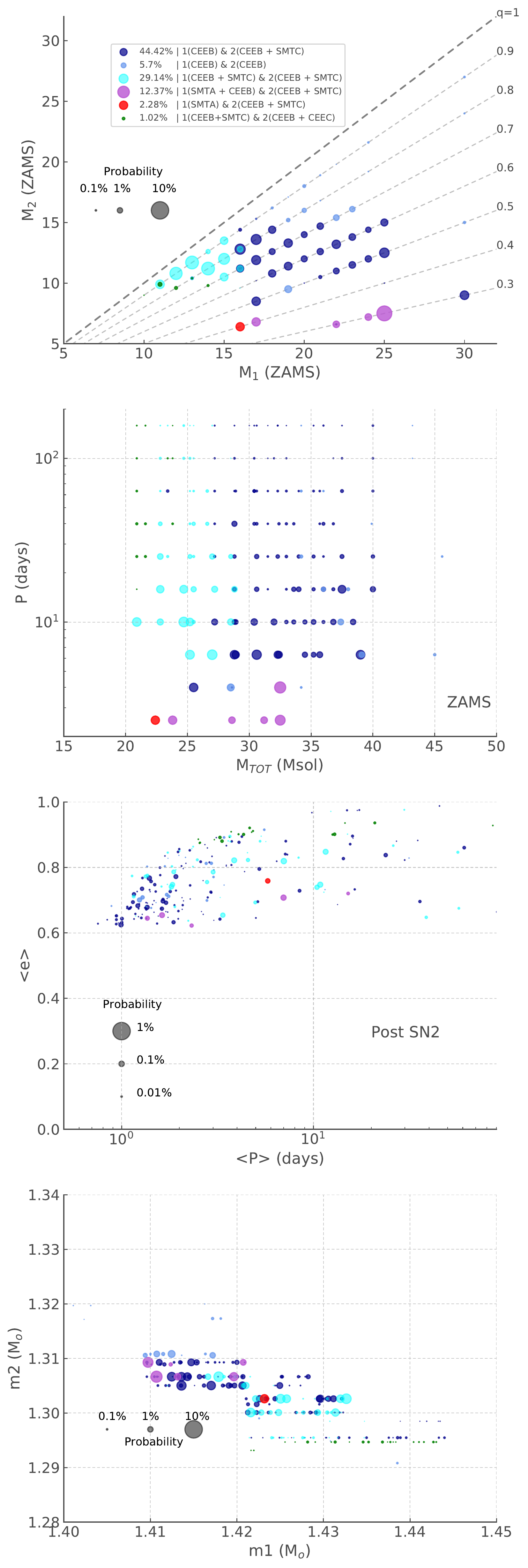}
    \caption{Same as Sup. Fig. \ref{fig:flowchartMPA1} for the AP 4 EoS}
    \label{fig:flowchartAP4}
\end{figure}

\begin{figure}[h!]
    \centering
    \includegraphics[width=8cm]{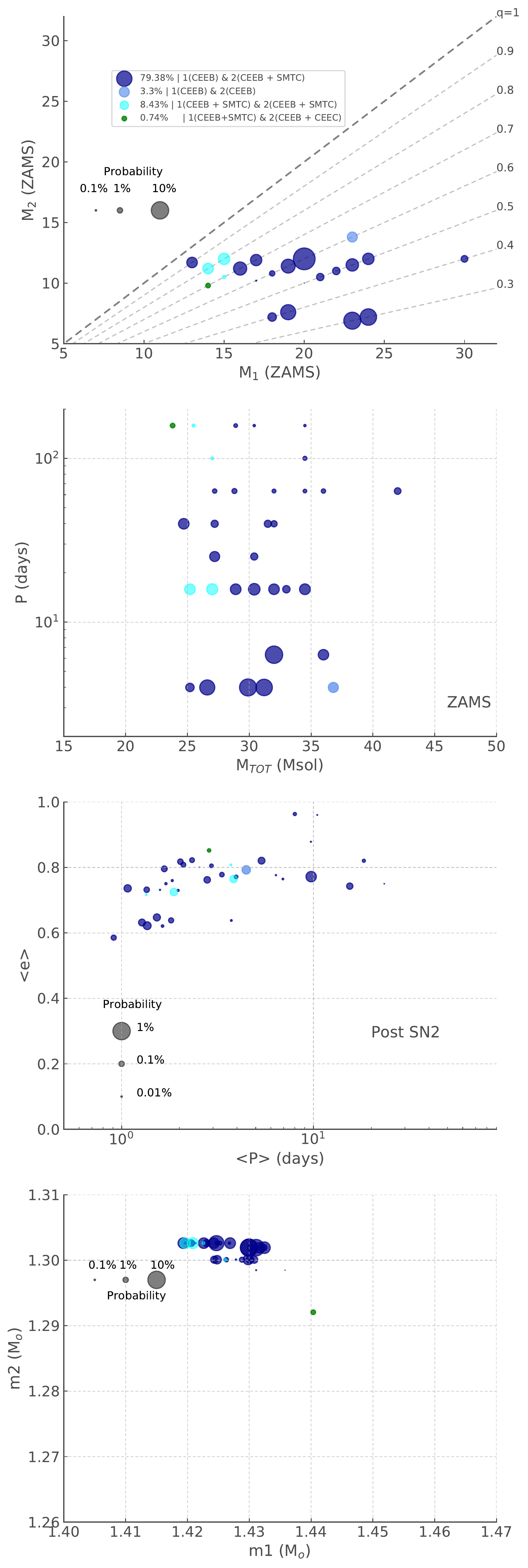}
    \includegraphics[width=8cm]{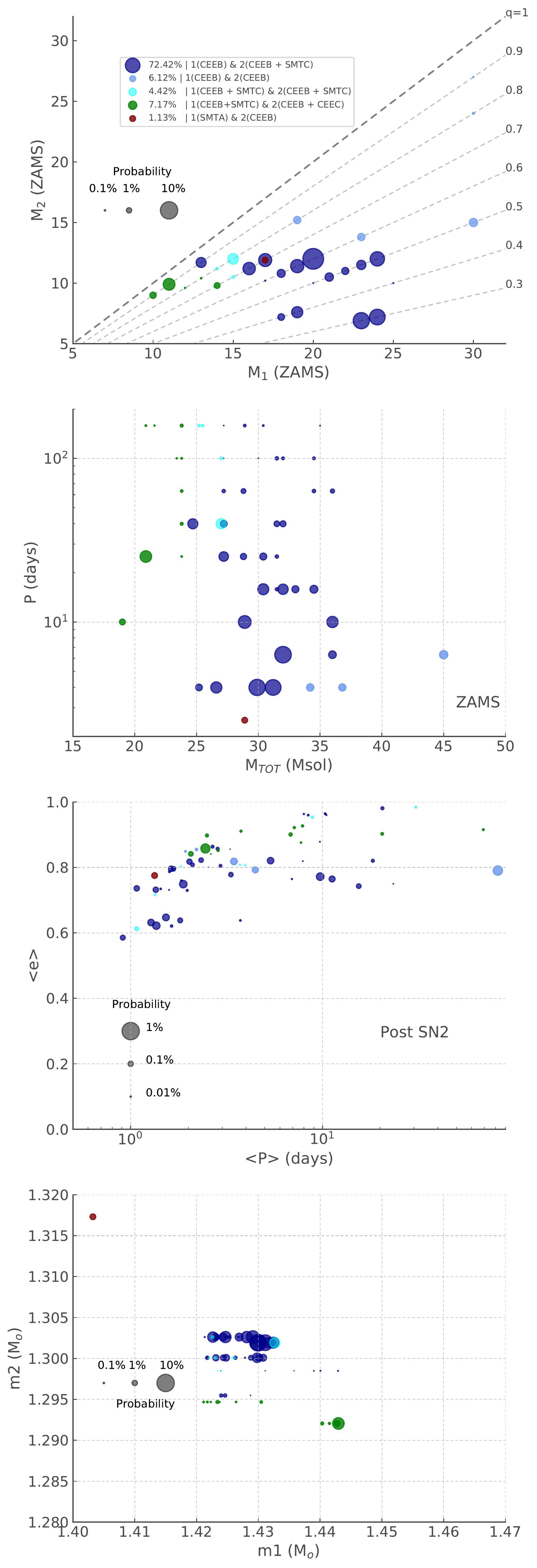}
    \caption{Same as Sup. Fig. \ref{fig:flowchartMPA1} for the WFF1 EoS}
    \label{fig:flowchartWFF1}
\end{figure}

%\subsection{MPA1 and AP4}

Across all of the genealogies retrieved CEE is a ubiquitous process, happening at least once in all systems - this is consistent with typical BNS channels (or other compact object binaries) described in the literature\cite{tauris2006, tauris2017, belczynski2018, vigna-gomez2018,chruslinska2018, olejak2021}.
A major difference between the BPASS results and the evolutionary channels found by the {\sc compas} and {\tt StarTrack} teams is that we find CEE to happen twice in most cases (where the primary star and the secondary star both go through their own CEE phase), and occasionally three times (where the secondary undergoes an additional CEE phase after He core ignition).
We explore this further in the next three sections. 

\subsection{Main progenitor pathways: case B CEE in the primary and in the secondary}
\label{sec:2CEE}
Channels with CEE in both the primary and the secondary star evolution have been seen in the {\tt StarTrack} code for BHBH/BHNS system formation\cite{olejak2021} and BNS formation\cite{chruslinska2018} where in the latter case double CEE channels are shown to arise in 48\% of cases (see their figure 1). 
We find that across the board most of our progenitor candidates undergo case B CEE without mass transfer occurring on the main sequence previously (see next section) -- in Sup. Fig. \ref{fig:flowchartMPA1}, \ref{fig:flowchartAP4} and \ref{fig:flowchartWFF1} they are denoted by the blue hues.

There are two main clusters amongst these channels.
The systems where the primary star also undergoes case C SMT (cyan) stand out as showing an upper mass limit for the primary \about 16 \msol and for the total mass of the system \about 30 \msol. 
Additionally all these systems have very high mass ratios with q greater than \about 0.7.
They also do not have initial periods below 6 days. 
Conversely the other channels with two case B CEE but no case C SMT in the primary have a wider range of permitted mass ratios (as low as 0.3) and although they prefer higher mass systems (total mass greater than \about 30 \msol and primary mass greater than \about 16\msol), there are a few exceptions, particularly in the systems with initial periods \about 4 days.

\subsection{Stable mass transfer on the main sequence}
\label{sec:1CEE}
There are two main types of channels where the primary star exhibit SMT on the main sequence (case A): those where the primary also goes through case B CEE (purple - two CEE over the life of the system) and those that do not (red hues - one CEE over the life of the system).
Both of these types of channels occur at low mass ratio (q \about 0.4 or lower) and can only happen if the initial period of the system is very short: typically \about 2.5 days (up to 4). Note that this is the lowest initial period in the BPASS stellar population grids so lower initial periods are likely possible but not covered by our parameter space.
The main difference between the systems that only undergo one CEE in the life of the secondary (red hues) and the others (purple) is that the former only occur for initial masses \about 14 -- 17 \msol whereas the latter cover a broader range of possible initial masses (up to 25 \msol when using the high spin prior). 

Overall we find that very few systems with evolutionary channels that undergo only one CEE ($<$5\%) across all the EoS and priors tested, which is at odds with results from the rapid population synthesis code {\sc compas} which predicts two main channels, neither of which undergo CEE twice\cite{vigna-gomez2018, vigna-gomez2020}.
This difference stems from a number of reasons: (i) the CEE prescription in {\sc compas} will lead to mergers more often than the detailed BPASS models due to their parameterisation\cite{compas2021}; (ii) their studies had the initial goal to reproduce the Galactic BNS population and it has been noted that this sample is not necessarily indicative of BNS systems in other galaxies, so a direct comparison to our GW170817 channels may not be completely appropriate \cite{pankow2018}; (iii) their Channel II (with mass rations \about 1 and both stars undergoing RLOF simultaneously) is actually not possible to obtain in BPASS because of our separation of the primary and secondary evolution (see Section \ref{sec:tui}). 
In population synthesis this very even mass ratio is extremely rare ($<1$\%), but if it is an effective way to create BNS systems its rate could be significantly boosted (~21\% in {\sc compas}\cite{vigna-gomez2020}). 
Alternatively it is possible that the difference in CEE treatement with the rapid population code and their potential propensity to overestimate the number of mergers, may require an exotic channel to explain the BNS rates observed in the Galaxy. 
It is not within the scope of this paper to draw a conclusion on this matter, but it highlights another consequence of the varied approached to CEE. 
These systems are unlikely to contribute significantly to the progenitor channels of GW170817-like events, or BNS merger rates in general, but it illustrates the variety of initial conditions that can converge towards the same evolution later in the life of the system (in this particular case, various types of systems can lead the the same secondary model). 
This is not unexpected, as it has previously been discussed that many initial conditions (such as mass ratio, separations and eccentricities) do not significantly affect merger rates\cite{mink2015}. 
This is important to note as this means that merger rates (or BNS rates) alone cannot be used easily to discriminate between evolutionary channels initial conditions. 

\subsection{Two CEE phases in the life of the secondary}
\label{sec:3CEE}
Some of our genealogies depart from the most well established channels in that the secondary star undergoes two phases of CEE (green hues).
This is more prevalent in the results found using the WFF1 EoS, although 1 to 2 \% of systems in the matches obtained with the AP4 EoS also follow this route, see Sup. Fig. \ref{fig:flowchartMPA1}, \ref{fig:flowchartAP4} and \ref{fig:flowchartWFF1}. 
These systems stand out from the previously described channels as having the lowest initial masses (with initial primary mass lower than \about 15 and total initial mass between 15 and 25\msol) and typically high mass ratios (q$>0.5$). The initial periods cover a wide range but notably the final periods and eccentricity are typically high, with e$>$0.8 and periods of a few days.

Interestingly the kick velocities (after the 2nd SN) recorded for these systems is significantly weighted towards lower velocities (see Sup. Fig. \ref{fig:3EoS_kick}) compared to a standard Hobbs distribution. 
This suggests that the kicks required for these genealogies to successfully create a GW170817-like progenitor system may have preferential magnitudes and/or orientations -- we discuss this further in Section \ref{sec:kicks}.

\begin{figure}[h!]
    \centering
    \includegraphics[width=10cm]{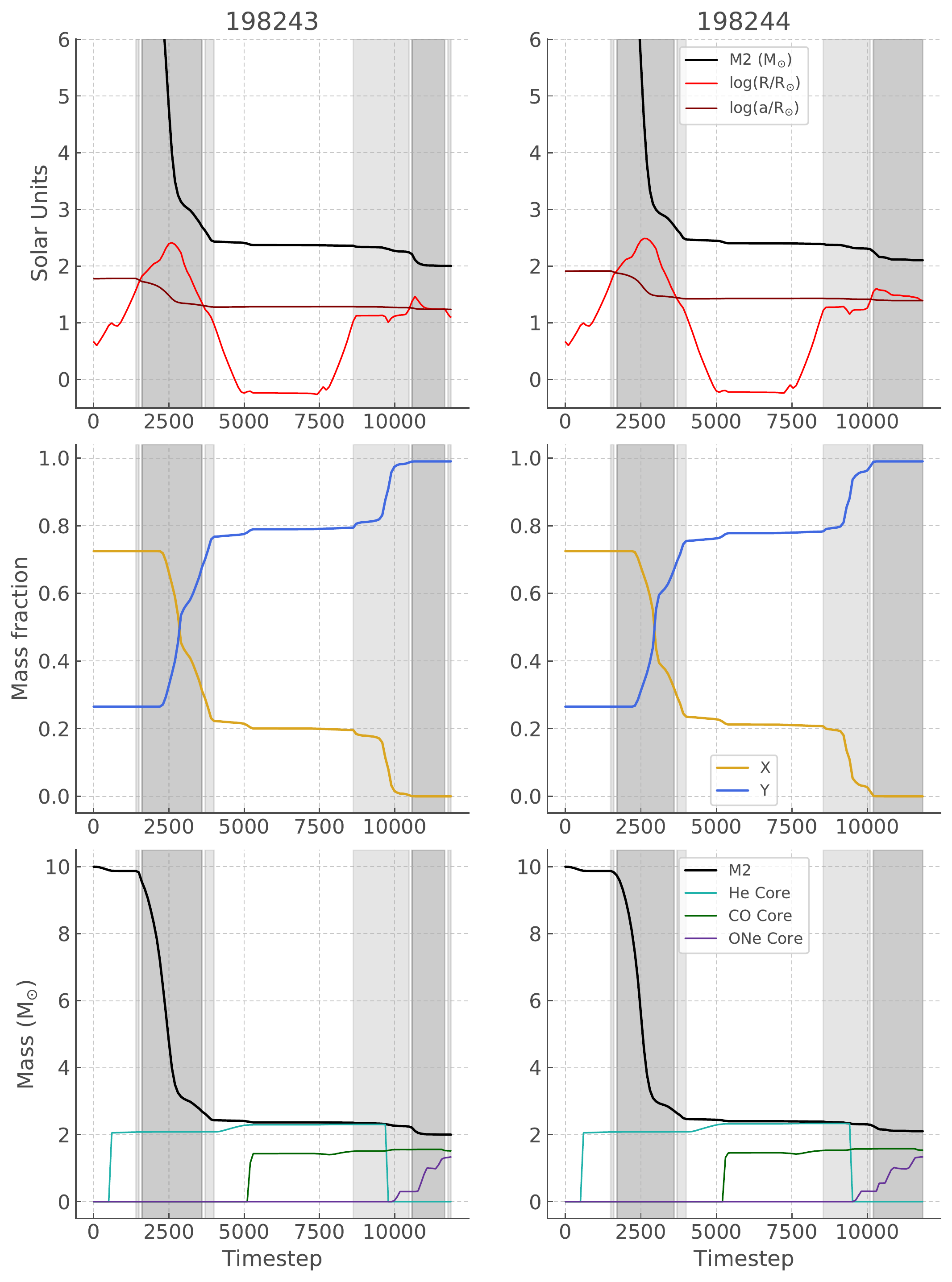}
    \includegraphics[width=10cm]{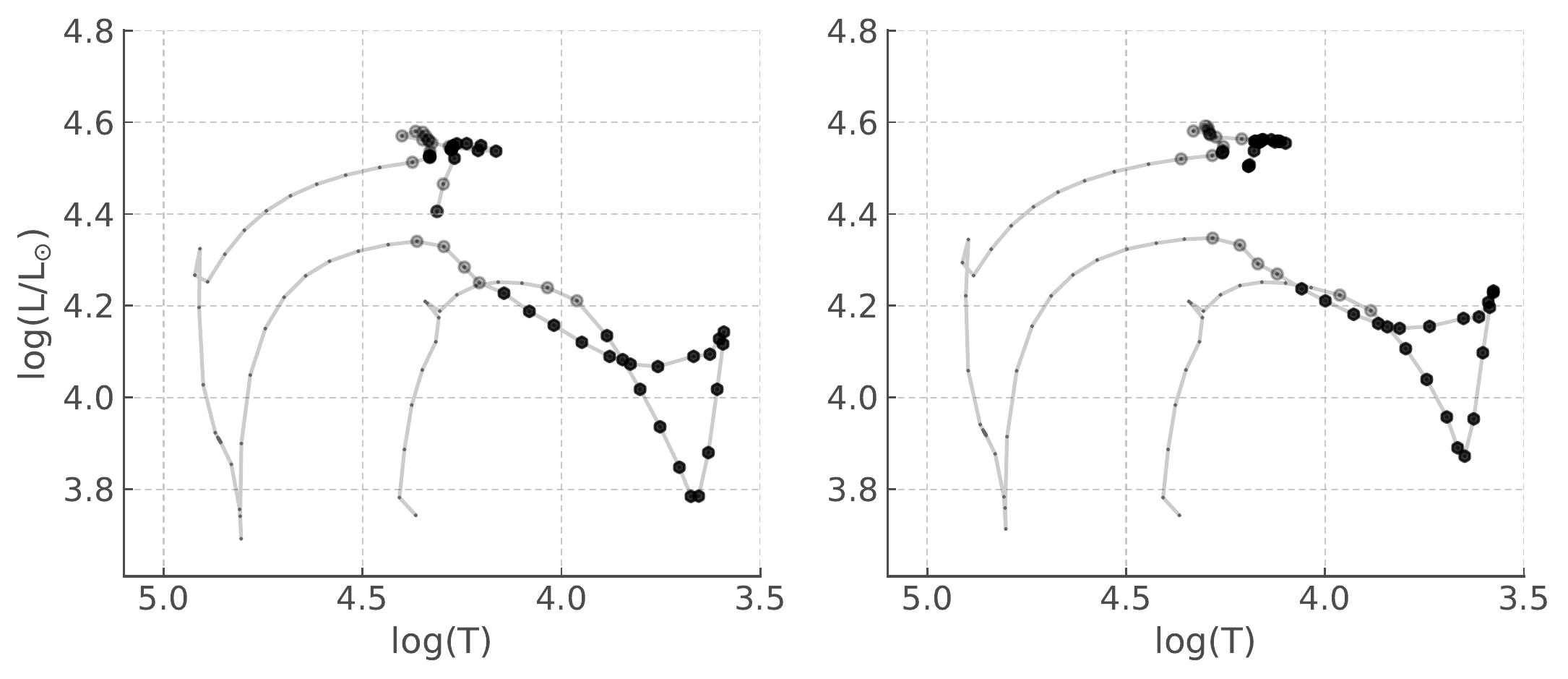}
    \caption{Evolution of the physical characteristics of two secondary models (ID numbers at the top) that exhibit two CEE phases - these are taken from the WFF1 high spin prior results. The dark shaded areas and black markers show the timesteps where the radius of the star is equal to or greater to the separation of the binary; the lighter shaded areas and grey markers show RLOF that does not satisfy our CEE condition. The timesteps are not evenly spaced in time - they are adaptively chosen by the stellar evolution code to capture significant changes in the evolution, such that short periods where the star undergoes significant, e.g. mass loss, will be sampled by more timestep than period of little change, e.g. during the main sequence. The use of the timesteps rather than age is deliberate so that it is easy to see the changes in the physical properties during the short, yet significant, phases of RLOF and CEE. Note that the X and Y abundances are surface abundances and the He core graph drops to 0 when the H shell is completely lost (X=0) as the He core mass coordinate becomes the full mass of the star.}
    \label{fig:CEEBC_plots}
\end{figure}

These types of channel could not occur in rapid population synthesis codes which remove all the envelope following a CEE event, and even in detailed simulations a different choice of the bifurcation point will dictate whether a star retains enough hydrogen envelope to expand once more.
Additionally, as highlighted by previous work\cite{laplace2020}, population synthesis codes based on Hurley et al 2000\cite{hurley2000} do not thoroughly take into account the structure of stripped stars and the effects of the remaining hydrogen, which could impact the re-occurance of a CEE envelope. 
We show in Sup. Fig. \ref{fig:CEEBC_plots} details of three stellar models which undergo two CEE phases. 
We find that the second CEE phase begins when the stars have hydrogen mass fractions X of \about 0.5 to 1.5\%, and this phase seems correlated with the onset of O-Ne burning in the core.
The scale of this CEE phase is also very different from the CEE seen after the onset of Helium burning - the expansion of the envelope does not extend very far relative to the separation, and its duration, mass loss and hardening of the orbit is found to be roughly an order of magnitude smaller across our models. 
%The short duration is likely a consequence of how easy it is for the tenuous remnant of envelope to be unbound. 

To further investigate the validity of these models with two CEE events, we search for them across all BPASS models: given that our secondary models have typically larger values of $\alpha \lambda$ (due to the effects of tides, see Section \ref{sec:CEE}), we want to ensure these phases are not just a numerical consequence of this. 
We find similar models in both primary models and secondary models, meaning that they are not the result of the apparent efficiency difference between CEE in our primary and secondary models.
We suggest that these late CEE phases following a previous CEE event can indeed be found in nature, and are simply not predicted by other models because of their assumptions (or because they end simulations before ONe burning). 
Moreover, this second CEE phase will be different from the first CEE event in the life of the star due to the tenuous envelope.
On the whole, such systems are rare - in the BPASS Z=0.010 population they are found in 2.5\% of model files corresponding to 0.06\% of systems (when weighted by $N_{\rm}(M_{1,\rm zams}, M_{2,\rm zams}, P_{\rm zams})$), but they may play a more significant role towards the creation of BNS systems (by a factor of 4 to 10 in the MPA1 and WFF1 results).

This is re-occurance of the CEE in the secondary model for BNS progenitors is not actually unique to BPASS: it has been reported from simulations done with the {\tt StarTrack} code and with the same metallicity\cite{chruslinska2018, belczynski2018}.
The key difference is that in these simulations the resulting system merges quickly after the second CEE phase as a result of a more significant orbital shrinkage: they find a separation coming out of the second CEE of order 0.1 \rsol (see their figure C1), which is 100 times smaller than what is seen in Sup. Fig. \ref{fig:CEEBC_plots}.
This is likely a direct consequence of our respective treatment of CEE, as our first CEE phase also results in less orbital shrinkage: our systems separation decreases by half an order of magnitude while theirs decreases by over 2 orders of magnitude.  

Overall it seems most likely that triple CEE evolutionary channels are able to account for a small fraction of BNS progenitors, but that the dominant route (at least within this comparison to GW170817) involved two CEE phases: one in the life of the primary and another in the life of the secondary.

%\begin{figure}[h!]
%    \centering
%    \subfloat[]{\includegraphics[width=6cm]{./figures/z010_AP4Pe_1CEE.png}}
%    \subfloat[]{ \includegraphics[width=6cm]{./figures/z010_AP4Pe_2CEE.png}}
%    \subfloat[]{ \includegraphics[width=6cm]{./figures/z010_AP4Pe_3CEE.png}}
%    \caption{Period and eccentricity of the new born BNS of the evolutionary channels found with AP 4 EoS}
%    \label{fig:AP4_Pe}
%\end{figure}

\subsection{Kick velocities after the second supernova}
\label{sec:kicks}

\begin{figure}[h!]
    \centering
    \includegraphics[width=17.5cm]{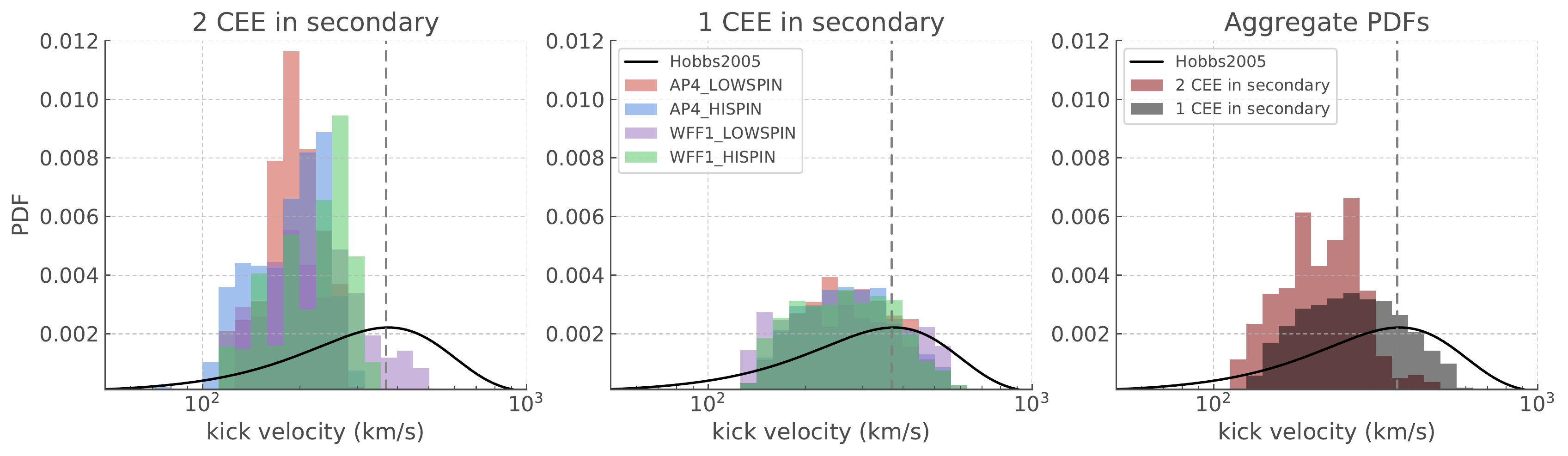}
    \caption{Kick velocity distributions of GW170817 progenitor channels. The posterior distribution of the kick velocities are grouped according to whether the secondary star undergoes one or two episodes of CEE. Note that the solutions for the MPA1 EoS are not shown because there is no instance of two CEE episodes in the life of the secondary stars.}
    \label{fig:3EoS_kick}
\end{figure}

So far in this work we have only considered a standard Hobbs kick distribution\cite{hobbs2005}, but previous studies looking to explain the Galactic BNS sample have called for lower kick velocities following the death of the secondary star as an Ultra-Stripped SN (USSN)\cite{tauris2013, tauris2015, tauris2017}. 
They arise from a star $\lesssim2$ \msol and result in very low ejecta-mass $\lesssim0.2-0.3$\msol\cite{de2018,yao2020}, which is responsible for the small kick values (of order tens of \kms rather than hundreds), with v$_{\rm kick}=30$\kms often being used in simulations\cite{compas2021}.

We show the kick distributions of the our matching genealogies for with the AP4 and WFF1 EoS in Figure \ref{fig:3EoS_kick}; we split the genealogies into two categories, the ones where the secondary star undergoes to epochs of CEE and the others. 
Firstly we note that on the whole the kick distributions of the matching progenitors are skewed to lower values than prior Hobbs distribution used. This is consistent with expectations as the Hobbs kick is based on the observation of runaway pulsars (which became unbound, as opposed to neutron star merger progenitors). 
The most skewed velocity distributions are seen for our channels where the secondary experiences a second episode of CE -- these also exhibited high period and eccentricity distributions at the birth of the BNS systems, suggesting they may require "special" kicks to be formed. 
Lower kick velocities, and potentially a specific direction (not yet recorded in our simulations, so we cannot quantify it at this stage) are therefore required for these channels to contribute to the progenitor BNS budget and we call these routes "lucky-kick" channels. 

As a sanity check we also performed numerical experiments including an USSN kick prescriptions for final masses $<2$\msol and found no match to our criteria for GW170817.
This could suggest that an USSN did not occur in the progenitor of GW170817 -- BNS can form and merge whilst being subjected to large kicks from the second SN\cite{zevin2020} -- or it could also be the result of the simplified kick prescription.
The value of v$_{\rm kick}=30$\kms has been found to work well for other populations synthesis codes studying compact mergers, but difference between their stellar evolution (especially CEE prescriptions) and that of BPASS may lead to different distributions of final periods.
A broader range of USSN kick values could provide some successful progenitor genealogies but they are unlikely to be a requirement for the progenitor of GW170817 as the kicks distributions of our successful models show high velocities.
In the context where the BPASS CEE results in less orbital shrinkage than other simulations, we would \textit{a priori} expect BPASS BNS progenitors to rely more heavily on USSNe, as lower kick velocities are required to avoid unbinding softer binaries. Therefore these results strongly support the idea that USSN and low kick velocities are not a requirement for all BNS merger pathways\cite{chu2021}.
This does not necessarily exclude the need for low kick velocities in the Galactic BNS sample.

Nevertheless, it seems that the Hobbs prescription is not entirely appropriate; in an upcoming study we will implement a broad range of kick prescriptions in TUI (including variations on the USSN kick velocity) and quantify their effects on BPASS compact object merger channels as a whole, we will also record the kick orientations as they may play a role in lucky-kick channels.